\documentclass[12pt,preprint]{aastex}
\newcommand{\beq}{\begin{equation}}
\newcommand{\eeq}{\end{equation}}
\def\stacksymbols #1#2#3#4{\def\theguybelow{#2}
        \def\verticalposition{\lower#3pt}
        \def\spacingwithinsymbol{\baselineskip0pt\lineskip#4pt}
        \mathrel{\mathpalette\intermediary#1}}
\def\intermediary #1#2{\verticalposition\vbox{\spacingwithinsymbol
        \everycr={}\tabskip0pt
        \halign{$\mathsurround0pt#1\hfil##\hfil$\crcr#2\crcr
                \theguybelow\crcr}}}

\catcode`\@=11
\def\gsim{\ifmmode{\mathrel{\mathpalette\@versim>}}
    \else{$\mathrel{\mathpalette\@versim>}$}\fi}
\def\lsim{\ifmmode{\mathrel{\mathpalette\@versim<}}
    \else{$\mathrel{\mathpalette\@versim<}$}\fi}
\def\@versim#1#2{\lower 2.9truept \vbox{\baselineskip 0pt \lineskip 
    0.5truept \ialign{$\m@th#1\hfil##\hfil$\crcr#2\crcr\sim\crcr}}}
\catcode`\@=12

\def\pd#1#2{\partial #1\over {\partial #2}}
\def\brem{bremsstrahlung$\;\,$}

\def\Lsun{L_{\odot}}
\def\Msun{M_{\odot}}
\def\kb{k_{\rm B}}
\def\mpr{m_{\rm p}}
\def\eps{\epsilon}
\def\epsII{\eps_{\rm II}}
\def\epsbhw{\eps_{\rm BHw}}
\def\epsopt{\eps_{\rm opt}}
\def\epsUV{\eps_{\rm UV}}
\def\tauII{\tau_{\rm II}}
\def\tauopt{\tau_{\rm opt}}
\def\tauUV{\tau_{\rm UV}}
\def\tform{\tau_{\rm form}}
\def\tdyn{\tau_{\rm dyn}}
\def\tcool{\tau_{\rm cool}}
\def\tjeans{\tau_{\rm Jeans}}
\def\trot{\tau_{\rm rot}}
\def\tlag{\tau_{\rm lag}}
\def\taul{\tau_{\rm *l}}
\def\tauh{\tau_{\rm *h}}
\def\lb{L_{\rm B}}
\def\lbh{L_{\rm BH}}
\def\lx{L_{\rm X}}
\def\lsn{L_{\rm SN}}

\def\ledd{L_{\rm Edd}}
\def\ldwin{L_{\rm dw}}
\def\luv{L_{\rm UV}}
\def\lopt{L_{\rm opt}}
\def\lbhefUV{L_{\rm BH,UV}^{\rm eff}}
\def\lbhefopt{L_{\rm BH,opt}^{\rm eff}}
\def\lbhefphot{L_{\rm BH,photo}^{\rm eff}}
\def\lbhef{\lbh^{\rm eff}}
\def\luveff{\luv^{\rm eff}}
\def\lopteff{\lopt^{\rm eff}}
\def\lir{L_{\rm IR}}
\def\lduv{L_{\rm d,UV}}
\def\ldopt{L_{\rm d,opt}}
\def\lrad{L_{\rm r}}
\def\mast{M_*}
\def\mhal{M_{\rm h}}

\def\Min{M_{\rm inf}}
\def\MII{M_{\rm II}}
\def\MTO{M_{\rm TO}}
\def\mbh{M_{\rm BH}}

\def\mg{m_{\rm g}}
\def\ms{m_*}
\def\msl{m_{\rm *l}}
\def\msh{m_{\rm *h}}
\def\mw{m_{\rm w}}
\def\mrem{m_{\rm rem}}
\def\mfid{m_{\rm fid}}
\def\medd{m_{\rm Edd}}

\def\RM{{\cal R}}
\def\ML{\Upsilon_*}
\def\mdot{\dot\mbh}
\def\dmin{\dot M_{1}}
\def\drhoII{\dot\rho_{\rm II}}
\def\als{\alpha_*}
\def\asn{\alpha_{\rm SN}}
\def\Rsn{R_{\rm SN}}
\def\rhos{\rho_*}

\def\re{R_{\rm e}}
\def\rs{r_*}
\def\rh{r_{\rm h}}
\def\rx{R_{\rm X}}
\def\kes{\kappa_{\rm es}}
\def\kUV{\kappa_{\rm UV}}
\def\kopt{\kappa_{\rm opt}}
\def\kpho{\kappa_{\rm photo}}
\def\kIR{\kappa_{\rm IR}}
\def\tx{T_{\rm X}}

\def\fh{f_{\rm h}}

\def\freml{f_{\rm rem,l}}
\def\fremh{f_{\rm rem,h}}
\def\etaII{\eta_{\rm SN}}
\def\etaD{\eta_{\rm d}}
\def\etaw{\eta_{\rm w}}
\def\etas{\eta_*}
\def\etaform{\eta_{\rm form}}

\def\Esn{E_{\rm SN}}
\def\NII{N_{\rm II}}
\def\dEII{\dot E_{\rm II}}
\def\dEI{\dot E_{\rm I}}

\def\dEopt{\dot E_{\rm opt}}
\def\dEUV{\dot E_{\rm UV}}
\def\Juveff{J_{\rm UV}^{\rm eff}}
\def\Jopteff{J_{\rm opt}^{\rm eff}}

\def\vw{v_{\rm w}}

\def\sigap{\sigma_{\rm a}}
\def\sigast{\sigma_*}

\def\drhosp{\dot\rhos^+}
\def\prad{p_{\rm rad}}
\def\t15{t_{15}}
\def\vtsn{\vartheta_{\rm SN}}

\slugcomment{Resubmitted version - May 10, 2007}

\lefthead{L. Ciotti, J.P. Ostriker}
\righthead{BH masses and galaxy evolution}

\begin{document}
\title{Radiative feedback from massive black holes
       in elliptical galaxies.\\
       AGN flaring and central starburst fueled by recycled gas}

\author{Luca Ciotti\altaffilmark{1} and 
Jeremiah P. Ostriker\altaffilmark{2,3}}
\affil{$^1$Department of Astronomy, University of Bologna,
via Ranzani 1, I-40127, Bologna, Italy} 
\affil{$^2$Princeton University Observatory, Princeton, USA}
\affil{$^3$IoA, Cambridge, UK}

\begin{abstract} 

The importance of the radiative feedback from massive black holes at the
centers of elliptical galaxies is not in doubt, given the well
established relations among electromagnetic output, black hole mass
and galaxy optical luminosity.  We show how this AGN radiative output
affects the hot ISM of an isolated elliptical galaxy with the aid of a
high-resolution hydrodynamical code, where the cooling and heating
functions include photoionization plus Compton heating.  We find that
radiative heating is a key factor in the self-regulated coevolution of
massive black holes and their host galaxies and that 1) the mass
accumulated by the central black hole is limited by feedback to the
range observed today, and 2) relaxation instabilities occur so that
duty cycles are small enough ($\lsim 0.03$) to account for the 
very small fraction of massive ellipticals observed to
be in the "on" -QSO- phase, when the accretion luminosity approaches
the Eddington luminosity.  The duty cycle of the hot bubbles inflated
at the galaxy center during major accretion episodes is of the order
of $\gsim 0.1-0.4$. Major accretion episodes caused by cooling flows in the
recycled gas produced by normal stellar evolution trigger nuclear
starbursts coincident with AGN flaring.  During such episodes the
central sources are often obscured; but overall, in the bursting
phase ($1\lsim z\lsim 3$), the duty-cycle of the black hole in its
"on" phase is of the order of percents and it is unobscured approximately
one-third of the time. Roughly half of the recycled gas from dying
stars is ejected as galactic winds, half is consumed in central
starbursts and less than 1\% is accreted onto the central black
hole. Mechanical energy output from non-relativistic gas winds
integrates to $2.3\times 10^{59}$ erg, with most of it caused by
broadline AGN outflows.  We predict the typical properties of the very
metal rich poststarburst central regions, and we show that the
resulting surface density profiles are well described by Sersic
profiles.
  
\end{abstract}


\section{Introduction}

Supermassive black holes (SMBHs) at the centers of bulges and
elliptical galaxies have played an important role in the processes of
galaxy formation and evolution (e.g., see Silk \& Rees 1998; Fabian
1999; Burkert \& Silk 2001; Cavaliere \& Vittorini 2002; King 2003;
Wyithe \& Loeb 2003; Haiman, Ciotti \& Ostriker 2004; Granato et
al. 2004; Sazonov et al. 2005; Murray, Quataert \& Thompson 2005; Di
Matteo, Springel \& Hernquist 2005; Begelman \& Nath 2005; Hopkins et
al. 2006; Croton et al. 2006), as strongly supported by the remarkable
correlations found between host galaxy properties and the masses of
their SMBHs (e.g., see Magorrian 1998, Ferrarese \& Merritt 2000,
Gebhardt et al. 2000, Yu \& Tremaine 2002, McLure \& Dunlop 2002,
Graham et al. 2003).

An additional very severe issue that AGN feedback likely addresses is
that of the "cooling flow problem" (e.g., Xu et al. 2002, Peterson \&
Fabian 2006), which in elliptical galaxies is at least as serious as
the heavily analyzed cooling flow problem in clusters (because the
radiative cooling times are an order of magnitude shorter, $\sim
10^{7.5}$ yr vs. $\sim 10^9$ yr). Also, a rarely discussed problem is
that the recycled gas (primarily from red giant winds and planetary
nebulae) of the evolving stellar population expected to accumulate
within galaxies over cosmic time, and responsible for the cooling
flow, is orders of magnitude larger than the mass of either the
central SMBH or the resident diffuse gas.  Quantitatively, the mass
return rate from evolving stars in elliptical galaxies can be
estimated as
\beq 
{\dot\mast}(t)\simeq 1.5\times {\lb\over 10^{11}L_{\rm B\odot}}
                     \t15^{-1.3} \quad \Msun {\rm yr}^{-1}, 
\eeq 
where $\lb$ is the present galaxy blue luminosity and $\t15$ is time
in 15 Gyr units (Ciotti et al.  1991, hereafter CDPR, see also
Sect. 2.2). Thus, the total mass return would accumulate onto the
central SMBH a mass by far too large compared with the observed SMBH
masses if a long lived cooling flow occurred.  Young stellar
populations observed in the body of ellipticals also cannot account
for the total mass released, and alternative forms of cold mass
disposal (such as distributed mass drop-out/star formation), are not
viable solutions (e.g., Binney 2001).  In addition to this mass
disposal problem, also the X-ray luminosity $\lx$ of low-redshift
elliptical galaxies is inconsistent with the standard cooling flow
model. In fact, low-redshift elliptical galaxies with optical
luminosity $\lb\gsim 3\times 10^{10}\Lsun$ show a significant range in
the ratio of gas-to-total mass at fixed $\lb$, with tabulated values
ranging from virtually zero up to few \% (e.g., Roberts et al. 1991),
and most of that is seen in X-rays with temperatures close to the
virial temperatures of the systems ($\sim 10^{6.7}\,$K, e.g.
O'Sullivan, Ponman \& Collins 2003).

A (partial) solution to these problems was proposed by D'Ercole et
al. (1989) and CDPR, by considering the effect of SNIa heating of the
galactic gas, and exploring the time evolution of gas flows by using
hydrodynamical numerical simulations. Subsequent, more realistic
galaxy models (with updated rates of SNIa derived by direct counts
from optical observations) were explored by Pellegrini \& Ciotti
(1998). It was found that while SNIa input sufficed for low and
medium-luminosity elliptical galaxies to produce fast galactic winds,
the inner parts of more massive spheroids would nevertheless host
inflow solutions similar to cooling flows.  This is because, while the
number of SNIa per unit optical luminosity is expected to be roughly
constant in ellipticals, the gas binding energy per unit mass
increases with galaxy luminosity.  However, it is not likely that
SNIa's can provide the entire solution to this problem, because the
distribution of the energy input is not concentrated enough to balance
the observed gas cooling rate (since this scales as $\rho^2$, the
required heating rate must be very large in the very central
regions). Note also that the SNIa rate is independent of the current
thermal state of the X-ray emitting plasma, therefore SN heating
cannot act as a self-regulating mechanism. Finally, as already
recognized by CDPR, the mass budget problem would still affect
medium-large galaxies, putative hosts of luminous cooling flows. Thus,
a concentrated feedback source is a very promising solution for a
variety of problems, and the central SMBH is the natural candidate, by
its mass and by its location, through a combination of mechanical and
radiative feedback mechanisms (e.g., see Fabian, Celotti \& Erlund
2006, and references therein).

Some calculations have allowed for a physically motivated AGN feedback
(e.g., see Binney \& Tabor 1995; Ciotti \& Ostriker 1997, 2001,
hereafter CO97, CO01; Omma et al. 2004; Ostriker \& Ciotti 2005;
hereafter OC05, Churazov et al. 2005), and the computed solutions are
characterized by relaxation oscillations (Ostriker et al. 1976; Cowie,
Ostriker \& Stark 1978).  Energy output (radiative or mechanical) from
the central SMBH pushes matter out, the accretion rate drops
precipitously and the expanding matter drives shocks into the galactic
gas. Then the resulting hot bubble ultimately cools radiatively (it is
thermally unstable) and the resulting infall leads to renewed
accretion; the cycle repeats.  Among the computed models which studied
the interaction between AGN feedback and galactic cooling flows, those
of CO97 and CO01 focused on the effects of {\it radiative heating} on
galactic gas flows. In fact, if one allows the radiation emitted from
the accreting SMBH to interact with and heat the galactic gas, one
solves the cooling flow problem in elliptical galaxies, and the
feedback produces systems that are variable but typically look like
normal ellipticals containing hot gas. They sometimes look like
incipient cooling flows and rarely, but importantly, appear like
quasars. Interestingly, observations seem to support this scenario
(e.g., Russell, Ponman, \& Sanderson 2007).

In CO97 and CO01, however, a major uncertainty remained about the
typical QSO spectrum to adopt, in particular the high energy component
of that spectrum, which is most important for heating the ambient gas.
Thus, a simple broken power law was adopted for the spectrum with a
range of possible values of the Compton temperature -- from
$10^{7.2}\,$K to $10^{9.5}\,$K -- with most of the emphasis of the
paper being on the higher temperatures.  Subsequent work by Sazonov,
Ostriker \& Sunyaev (2004, see also Sazonov et al. 2007), which
assessed the full range of observational data of AGNs and computed
their Spectral Energy Distribution, concluded that the typical
equilibrium radiation temperature was narrowly bounded to values near
$10^{7.3}\,$K, i.e., of the order of 2keV. This value, although it is
at the lower end of the range adopted by C001, is still above the
virial temperature of all galaxies and, most importantly, well above
the central temperature of the cooling flow gas.  As noted by Sazonov
et al. (2005), there is a rather large compensating effect also not
included in CO01: gas heated by radiation with a characteristic
temperature near $10^7\,$K is heated far more effectively by
absorption in the atomic lines of the abundant metal species than by
the Compton process. In particular, Sazonov et al. (2005) provide a
fitting formula for the Compton plus photoionization and line
heating/cooling that we implemented into our numerical code, together
with additional physics that was missing in CO97 and CO01; a very
preliminary exploration of the new models can be found in
OC05. Consistently with a second generation of metal rich stars
observed in recent SDSS surveys (e.g., Fukugita et al. 2004; Nolan,
Raychaudhury, \& Kab\'an 2007), we now also allow for star formation,
which is found of primary importance during the first few Gyr, as it
consumes a large fraction of the cooling gas in a central starburst
(e.g., Reuland et al. 2007).

In this paper we address the coevolution of a SMBH and of the ISM of
the host galaxy, and argue that the AGN/starburst feedback effects can
be the answer to the triple problems of (1) the suppression of cooling
flows within galaxies, (2) the large scatter in their hot gas mass at
fixed optical luminosity, and (3) disposition of the recycled gas. It
is found that the intermittencies described in CO97 and CO01 are
confirmed also with the more accurate treatment of the radiation
field, and these have two consequences: they cause nuclear starbursts
and they feed the central SMBH in what we observe as coincident
AGN/starburst episodes. In addition, since the fuel is in fact
proportional to the evolving stellar mass, one maintains - on average
- a central SMBH mass and a younger stellar population mass that are
both proportional to the galactic stellar mass, with the energetic
feedback from central SNII and AGN producing the energy input needed
to keep the bulk of the gas in a state of low density and high
temperature.  For simplicity in this paper we just present a summary
of the main properties of a typical model, leaving to following papers
the detailed description of specific features of observational and
theoretical interest.

The paper is organized as follows. In Section 2 we describe how the
galaxy models adopted in the simulations are built, the details of the
input physics, and their numerical implementation. In Section 3 the time
evolution of a representative model galaxy is illustrated in detail.
Finally, in Section 4 we discuss the main results obtained.

\section{The models}

The galaxy models and the input physics adopted for the simulations
have been substantially improved with respect to CO97 and CO01: in the
following we describe them in detail for future reference.

\subsection{Structure and internal dynamics}

In CO01 the galaxy models utilized a King (1972) stellar distribution
plus a quasi-isothermal dark matter halo, in line with the models then
used for cooling-flow studies. However, the existence of large cores
of constant surface brightness has been clearly ruled out (e.g.,
see Jaffe et al. 1994, Faber et al. 1997, Lauer et al. 2005), as HST
observations have shown how the central surface brightness profile is
described by a power-law as far in as it can be observed, i.e. to
$\sim 10$ pc for Virgo ellipticals. Following Pellegrini \& Ciotti
(1998), we then adopt a stellar density distribution described by the
more appropriate Hernquist (1990) model
\beq
\rhos = {\mast\over 2\pi}{\rs\over r(r+\rs)^3}.
\eeq
Optical (e.g., Saglia et al. 1993; Bertin et al. 1994; Cappellari et
al. 2006, Douglas et al. 2007) and X-ray (e.g. Fabian et al. 1986,
Humprey et al. 2006) based studies typically find that luminous matter
dominates the mass distribution inside the effective radius $\re$,
while dark matter begins to be dynamically important at $2-3\re$, with
common values of the total dark-to-luminous mass ratio $\RM\equiv
\mhal/\mast$ in the range $1\lsim\RM\lsim 6$.  Theoretical (e.g.,
Dubinski \& Carlberg 1991; Ciotti \& Pellegrini 1992; Navarro, Frenk
\& White 1997; Fukushige \& Makino 1997) and observational (e.g., Treu
et al. 2006) arguments support the idea that similarly to the stellar
profiles, also the radial density distribution of the dark halos is
described by a cuspy profile with central spatial density increasing
as $r^{-1}$: consistently, for the dark matter halo we also adopt the
density distribution in equation (2), where $\mhal=\RM\mast$ and
$\rh=\beta\rs$ are the halo total mass and scale-length; dynamical and
phase-space properties of the resulting two-component Hernquist models
are given in Ciotti (1996).

In order to use realistic galaxy models, the mass distribution is
determined as follows. We fix a value for the projected central
velocity dispersion $\sigma$, and we determine the galaxy {\it present day}
model blue luminosity $\lb$ and effective radius $\re$ from the
Faber-Jakson
\beq
{\lb\over 10^{11}L_{\rm B\odot}}=
                        0.23\;\left(\sigma\over 300 {\rm km /s}\right)^{2.4} +
                        0.62\; \left(\sigma\over 300 {\rm km /s}\right)^{4.2}
\eeq
and  the Fundamental Plane
\beq
\log\re = A\log\sigma + B\log I_{\rm e}+ C
\eeq
relations adopted in CDPR, where $I_{\rm e}\equiv\lb/(2\pi\re^2)$. The
scale-length $\rs$ of the stellar distribution (2) is then given by
$\rs\simeq\re/1.8153$ (Hernquist 1990).

Having fixed $\sigma$ and $\rs$, we determine the galaxy stellar
mass and halo properties such that $\sigap$, the model aperture velocity
dispersion within $\re/8$ (obtained by solving and projecting the
Jeans equations), coincides with $\sigma$.  For globally isotropic
two-component Hernquist models
\beq 
\sigap^2\left({\re\over 8}\right)\simeq {G\mast\over\rs}
\left(0.096+{0.12\;\RM\over \beta^{1.72}}\right),
\eeq 
where 
\beq
\beta\equiv {\rh\over\rs}=(1+\sqrt{2})
            \left(\sqrt{2\;\RM\over\RM_{0.5}}-1\right),
\eeq 
and $\RM_{0.5}$ is the dark-to-visible mass ratio within the stellar
half-mass radius (Pellegrini \& Ciotti 2006); note that we do not
consider the effect of $\mbh$ on $\sigap$, in accordance with
estimates of the SMBH sphere of influence radius (e.g., Riciputi et
al. 2005).  For chosen values of $\RM$ and $\RM_{0.5}$ we obtain
$\mast$ and the stellar mass-to-light ratio $\ML\equiv \mast/\lb$ from
equations (5)-(6). In the initial conditions we then expand the scale
radius $\rs$ thus determined by a factor 1.5 (while maintaining fixed
the dark matter halo distribution and total mass), in order to allow
for the subsequent shrinkage of newly formed stars in the galaxy
central regions (see Sect. 2.3). As discussed in the Conclusions, we
do not attempt at this stage to properly follow the dark matter halo
contraction, nor the modifications of the velocity dispersion profile
consequent to star formation.

An important ingredient in the energetics of the gas flows, namely the
thermalization energy of the stellar mass losses, depends on the
radial trend of the stellar velocity dispersion (see Sect. 2.7) which,
for the isotropic case is given (via the Jeans equations) by
\begin{eqnarray}
\rhos\sigast^2 =\rhos\sigast^2 (\mbh=0)+
                {G\mast\mbh\over 2\pi\rs^4}
                    \left[6\ln\left(1+{1\over s}\right)
                          -{(1+2s)(6s^2+6s-1)\over 2s^2 (1+s)^2}\right],
\label{jeans}
\end{eqnarray}
where $s\equiv r/\rs$, and the first term at the r.h.s. is the
1-dimensional stellar velocity dispersion without the contribution of
the SMBH (Ciotti, Lanzoni \& Renzini 1996).

\subsection{Stellar passive evolution: SNIa rate and stellar mass losses}

The stellar mass loss rate and the SNIa rate associated with the
initial stellar distribution are the main ingredients driving
evolution of the models. In the code the stellar mass losses -- the
source of {\it fuel} for the activity of the SMBH -- follow the detailed
prescriptions of the stellar evolution theory, while for quick
calculations the approximation given in equation (1) can be used.
Over the whole galaxy
\beq
\dot\mast={\rm IMF}(\MTO)|\dot\MTO|\Delta M,
\eeq
where the initial mass function IMF is a Salpeter law (normalized as
described in CDPR), and the turn-off mass (in $\Msun$) of stars at time 
$t$ (in Gyrs) is
\beq
\log \MTO=0.0558(\log t)^2-1.338\log t +7.764.
\eeq
Finally
\beq
\Delta M=\cases{\MTO-M_{\rm fin}(\MTO)=0.945\MTO-0.503,\quad (\MTO< 9\Msun),\cr
                \Delta M=\MTO-1.4\Msun,\quad (\MTO\geq 9\Msun),}
\eeq
(CDPR).  Recently, updated formulas have become available (Maraston
2005), but in the present context the modifications on the flow
evolution are minor (Pellegrini \& Ciotti 2006) and so for continuity
we maintained the treatment of past works.  Also, SNIa rates in
early-type galaxies have been carefully reanalyzed, and current
estimates of the rate in the local universe agree on $0.32h^2$ SNu
(where 1 SNu = 1 SNIa per century per $10^{10}L_{\rm B\odot}$ and
$h\equiv H_{\circ}/100$ km s$^{-1}$ Mpc$^{-1}$; e.g., see Cappellaro,
Evans \& Turatto 1999; Mannucci et al. 2005) so that, following CDPR,
we parametrize the time evolution of the SNIa rate as
\beq
\Rsn (t)=0.32\times 10^{-12}h^2\vtsn {\lb\over L_{\rm B\odot}}
          \left ({t\over 13.7\,{\rm Gyr}}\right)^{-s}\quad {\rm yr}^{-1},
\eeq
where the coefficient $\vtsn$ allow for different choices in the
present-day SNIa.  Assuming for each supernova event an energy release of
$\Esn=10^{51}$ erg, a fraction $\etaII$ of which is thermalized in the
surrounding ISM, the energy input per unit time over all the galaxy
body is given by
\beq
\lsn (t)=1.015\times 10^{31}h^2\vtsn\etaII {\lb\over L_{\rm B\odot}}
          \left ({t\over 13.7\,{\rm Gyr}}\right)^{-s}
          \quad\quad {\rm erg}\,{\rm s}^{-1};  
\eeq
in this paper we restrict to the case $\vtsn =1$ and $h=0.75$.  As is
well known, the specific value of $s$ is a critical ingredient in the
model evolution. In fact, when $s\gsim 1.3$ the flow evolves from wind
to inflow, while the opposite is true for $s\lsim 1.3$. This was
especially true in CDPR, CO97 and CO01 models. However, the specific
value of $s$ is less important in affecting the evolution of gas flows
in cuspier galaxy models with a somewhat reduced SNIa present-day
rate, as those here explored, that preferentially host "partial wind"
solutions (Pellegrini \& Ciotti 1998). Thus, here we restrict to the
currently favoured $s=1.1$ value (Greggio 2005), even though a more
complicate time dependence than a simple power-law seems possible
(e.g., Matteucci et al. 2006; Neill et al. 2007).

Besides energy, supernovae provide also mass. We assume that each SNIa
ejects $1.4\Msun$ of material in the ISM, so that the total rate of
mass return from the aging initial stellar population at each place in
the galaxy is
\beq
{d\rhos\over dt}=(\als +\asn)\rhos ,
\label{alphas}
\eeq
where $\asn (t)=1.4\Msun\,\Rsn (t)/\mast$ and $\als (t)=\dot\mast (t)
/\mast$ are the specific mass return rates.  With these definitions,
the SNIa kinetic energy injection per unit volume in the ISM can be
written as
\beq
\dEI=\etaII\Esn{\Rsn\over\mast}\rhos=\etaII\Esn{\asn(t)\rhos\over 1.4\Msun}.
\label{dEI}
\eeq

\subsection{Star formation, SNII heating and starburst properties}

At variance with CO01, in the present study we allow for star
formation, which cannot be avoided when cool gas accumulates in the
central regions of elliptical galaxies. In particular, we compute the
star formation rate at each radius $r$ from the equation
\beq
\drhosp={\etaform\rho\over\tform},\quad
\tform=\max (\tcool,\tdyn),
\label{drhosp}
\eeq
where $\rho$ is the local gas density, $\etaform=0.03\div 0.4$ (e.g.,
see Weinberg, Hernquist, \& Katz, 2002; Cen \& Ostriker 2006), 
and the associated caracteristic times are
\beq
\tcool\equiv {E\over C},\quad
\tdyn=\min(\tjeans,\trot),\quad
\tjeans\equiv\sqrt{3\over 32\pi G\rho},\quad
\trot\equiv{2\pi r\over v_c(r)}.
\eeq
This formulation is usually termed the "Schmidt-Kennicutt
prescription". $E$ and $C$ are the gas internal energy and the
effective cooling per unit volume (see equation~[\ref{dotE}]), while
$v_c(r)$ is the galaxy rotational velocity at radius $r$. In the code
the stars are maintained in the place where they form, and in each
shell the associated sinks of momentum and internal energy per unit
volume are given by the negative of
\beq
\dot m_*^+={\etaform m\over\tform}, \quad 
\dot E_*^+={\etaform E\over\tform},
\eeq
where $m$ is the specific momentum of the ISM (see Sect. 2.7).

For a total mass $\Delta\mast$ of newly formed stars in a given
time-step and at a given place we assume a Salpeter IMF
\beq
{dN\over dM}=
(x-1)\left({\Min\over\Msun}\right)^{x-1}{\Delta\mast\over\Msun}\times
\left({M\over\Msun}\right)^{-1-x},\quad 
(x >1, M\geq\Min =0.1\Msun),
\label{dNdM}
\eeq
so that the associated total number of Type II Supernovae is
\beq
\NII=\int_{\MII=8\Msun}^{\infty}{dN\over dM}dM=\left(1-{1\over x}\right)
\left({\Min\over\MII}\right)^x{\Msun\over\Min}{\Delta\mast\over\Msun}
\simeq 7\times 10^{-3}{\Delta\mast\over\Msun},
\eeq
where the numerical value holds for $x=1.35$.  As for SNIa, we assume
that each SNII event releases $\Esn=10^{51}$ erg of kinetic energy, 
and the resulting mean efficiency is
\beq
\epsII\equiv {\NII\Esn\etaII\over\Delta\mast c^2} = 
\left(1-{1\over x}\right)
\left({\Min\over\MII}\right)^x{\Msun\over\Min}{\Esn\etaII\over\Msun c^2}
\simeq 3.9\times 10^{-6}\etaII;
\label{epsII}
\eeq
in this paper we assume $\etaII=0.85$.  The characteristic time for
SNII explosion is fixed to $\tauII=2\times 10^7$ yr, and from
equations~(\ref{drhosp}) and (\ref{epsII}) their luminosity (per unit
volume) at each radius from the galaxy center is
\beq
\dEII (t)\equiv {\epsII c^2\over\tauII}\int_0^t\drhosp (t')
e^{-(t-t')/\tauII}dt'.
\label{dEII}
\eeq

We assume that each explosion leaves a neutron stars of $1.4\Msun$;
another possibility would be to assume more massive BH remnants (e.g.,
see Renzini \& Ciotti 1993). As a consequence, the total mass ejected
by the SNII explosions per unit mass is
\beq
{\MII^{ej}\over\Delta\mast}=\left({\Min\over\MII}\right)^{x-1} 
                       -1.4{\NII\Msun\over\Delta\mast}\simeq 0.2,
\eeq
and the mass return rate per unit volume of the young evolving stellar
population is given by
\beq
\drhoII(t)\simeq {0.2\over\tauII}\int_0^t\drhosp (t')
e^{-(t-t')/\tauII}dt'.
\label{drhoII}
\eeq

Finally, in the code we also compute the fiducial optical and UV
luminosity per unit volume of the new stars as
\beq
\dEopt (t)\equiv {\epsopt c^2\over\tauopt}\int_0^t\drhosp (t')
e^{-(t-t')/\tauopt}dt',
\label{dEopt} 
\eeq
and 
\beq
\dEUV (t)\equiv {\epsUV c^2\over\tauUV}\int_0^t\drhosp (t')
e^{-(t-t')/\tauUV}dt',
\label{dEUV} 
\eeq
respectively, where $\epsopt =1.24\times 10^{-3}$, $\epsUV =8.65\times
10^{-5}$, $\tauopt =1.54\times 10^8$ yr, and $\tauUV =2.57\times 10^6$
yr are the efficiency and characteristic time of optical and UV
emission, respectively.  The time-delay equations~(\ref{dEII}) and
(\ref{drhoII})-(\ref{dEUV}) are integrated according to the scheme
described in Appendix.

\subsection{The circumnuclear disk and the SMBH accretion luminosity}

At the onset of the cooling catastrophe a large amount of gas suddenly
flows onto the central regions of the galaxy, and this induces star
formation and accretion on the central SMBH, producing a burst of
energy from the galaxy center.  However observations of our own
galactic center and high resolution studies of other nearby systems
indicate that, in addition to the central starburst with radius $\sim
100 - 300$ pc, accretion onto the central SMBH is mediated by a small
central disc within which additional significant star formation occurs
(e.g., see Goodman \& Tan 2004; Tan \& Blackman 2005), and the
remaining fraction of gas either is blown out in a broadline wind, or
it is accreted onto the central SMBH.  In our code the disk is not
simulated with hydrodynamical equations, but its modelization is
needed to obtain important quantities required by the code.  The disk,
which is the repository of the gas inflowing at a rate $\dot M_1$ from
the first active grid point $R_1$ of the hydrodynamical
grid\footnote{$\dmin$ is taken positive in case of accretion and zero
in case of outflow at $R_1$.}, and which feeds the central SMBH at a
rate $\dot\mbh$, contains at any time the mass gas $\mg$ and a total
stellar mass $\ms =\msl +\msh$, which is divided among low and high
mass stars.  Finally, the disk also contains a mass $\mrem$ of
remnants from the earlier generations of evolved stars.

In the adopted scheme the accretion rate on the central SMBH is given
by
\beq
\dot\mbh ={\dot\mfid\over 1+\etaD},
\eeq
where
\beq
\dot\mfid\equiv {\mg\over\tlag},\quad \dot\medd\equiv{\ledd\over\eps c^2},
                                \quad\etaD\equiv {\dot\mfid\over 2\dot\medd},
\eeq 
are the fiducial depletion rate of gas from the circumnuclear disk and
the Eddington mass accretion rate, respectively, while the
characteristic disk star-formation time $\tlag$ is defined as
\beq
\tlag\equiv{2\pi\over\alpha}\sqrt{R_1^3\over G\mbh}, 
\eeq 
where $\alpha\simeq 10^{-2}-10^{-1}$ (or higher, e.g., see King,
Pringle \& Livio 2007) is the disk viscosity coefficient.  More
rigorous formulations of $\tlag$ are possible, but at the current
level of modelization equation (28) is accurate enough. Thus,
equations (26)-(27) guarantee that when $\etaD <<1$ the gas is
accreted onto the central SMBH at the rate $\dot\mfid$, while
$\dot\mbh=2\dot\medd$ for $\etaD>>1$ (the factor of 1/2 in the
definition of $\etaD$ allowing for possible super-Eddington
accretion); note however that outside the first grid point $R_1$ the
flow accretion rate is limited in a self-consistent way by radiation
pressure (see Sects. 2.6 and 2.7).

We also assume that a fraction of the disk gas mass is converted into
stars at a rate $\etas\dot\mfid$ (where $\etas\simeq 10\mg/\mbh$, to
approximate a Toomre criterion for star formation in the disk), and
that another fraction of $\mg$ is lost as a wind at an instantaneous
rate given by $\etaw\dot\mbh$ (with $\etaw=2$, so that the broadline
wind carries away twice as much mass as falls to the SMBH), so that
the equation for the gas mass in the disk is
\beq
{d\mg\over dt}=\dmin - (1+\etaw)\dot\mbh -\etas\dot\mfid.
\eeq 

The stars formed in the disk are described separately as a function of 
their mass, i.e., high-mass stars ($M>\MII =8\Msun$) produce a total 
disk mass $\msh$, and  low-mass stars ($\Min <M<\MII$)
contribute to a disk mass $\msl$ according to the equations 
\beq
{d\msl\over dt}=(1-\fh)\etas\dot\mfid -{\msl\over\taul};\quad
{d\msh\over dt}=\fh\etas\dot\mfid -{\msh\over\tauh},
\eeq
where for the caracteristic evolutionary times we adopt $\taul
=\tauopt$ and $\tauh=\tauII$ given in Sect. 2.3, while we assume $\fh
= 0.5$, corresponding to a top-heavy Salpeter-like initial mass
function of slope $x\simeq 1.16$ and minimum mass $\Min = 0.1\Msun$
(see equation~[\ref{dNdM}])\footnote{The choice of a top-heavy
IMF for the disk stars, at variance with the standard Salpeter law
adopted for the new stars formed over the galaxy body, is motivated,
for example, by the dynamical and X-ray evidences reported in
Nayakshin \& Sunyaev 2005 and Nayakshin et al. 2006).}.  The associated
optical ($\ldopt$) and UV ($\lduv$) luminosities of the stellar disk
are calculated following the scheme of Sect. 2.3.  Finally stellar
remnants mass in the disk evolves as
\beq
{d\mrem\over dt}=\freml{\msl\over\taul}+\fremh{\msh\over\tauh},
\eeq
where $\freml=0.2$, $\fremh=0.09$. Thus, the equation for the mass of the 
disk wind is 
\beq
{d\mw\over dt}=\etaw\dot\mbh+
               (1-\freml){\msl\over\taul}+
               (1-\fremh){\msh\over\tauh}:
\label{eqmdiskw}
\eeq
the first term is a mass loss driven by the central SMBH and the
second and third are from high mass and low mass stars in the central
disk. All the equations in this Section are integrated with a first
order finite difference scheme.

From equation (26) we calculate the electromagnetic bolometric 
accretion luminosity as
\beq
\lbh(t) =\eps\,\mdot(t)\,c^2,
\label{lbhbol}
\eeq
where the radiative efficiency $\eps$ usually spans the range
$0.001\lsim\eps\lsim 0.15$.  We adopt $\eps=0.1$ in the present paper,
consistent with observations (e.g., see Soltan 1982, Yu \& Tremaine
2002, Haiman et al. 2004), but a generalization to include an
ADAF-like efficiency could also be easily implemented in the code
(Narayan \& Yi 1994, see also CO01).

We also compute a fiducial mechanical luminosity for the disk wind
$\ldwin$, defined as
\beq
\ldwin =\epsbhw\dot\mbh c^2 +\epsII c^2 (1-\fremh){\msh\over\tauh},
\label{eqldwin}
\eeq
where $\epsbhw=5\times 10^{-4}=0.005\eps$ is the mechanical efficiency
of the SMBH (Elvis 2006).  This mechanical energy output is a factor
of 10 lower than adopted by Hernquist and collaborators when described
in the same units, and adding this mechanical output to our simulation
would be only a slight correction, since $\epsbhw/\eps\sim 0.5$\%.
The associated disk wind velocity is then given by
\beq
\vw\equiv\sqrt{2\ldwin\over\dot\mw}\simeq\sqrt{2\epsbhw\over\etaw}c
    \simeq 7\times 10^3\;{\rm km}\; {\rm s}^{-1}
\eeq
for our parameters, in agreement with observations of broad line
regions.

In summary, the mass falling to the center is mediated by a gaseous, 
starforming $\alpha$-disk which surrounds the SMBH. A small
fraction of the mass is turned into (primarily) high mass stars in the
disk, roughly two thirds is expelled in the broad line wind and
roughly one third is accreted onto the central SMBH. 

\subsection{Radiative heating and cooling}

With an improvement over CO01, the radiative heating and cooling
produced by the accretion luminosity are numerically computed by using
the Sazonov et al. (2005) formulae, which describe the net
heating/cooling rate per unit volume $\dot{E}$ of a cosmic plasma in
photoionization equilibrium with a radiation field characterized by
the average quasar Spectral Energy Distribution by Sazonov et
al. (2004), whose associated spectral temperature is $\tx\simeq 2$
keV.  In particular, Compton heating and cooling, \brem losses, line
and recombination continuum heating and cooling, are taken into
account.

A good approximation to the net gas energy change rate $\dot E$, valid
for $T\gsim 10^4$ K (all quantities are expressed in cgs system) is
given by
\beq
\dot E = n^2 (S_1 + S_2 + S_3)\equiv H-C,
\label{dotE}
\eeq
where $n$ is the Hydrogen density (in number), and positive and
negative terms are grouped together in the heating ($H$) and cooling
($C$) functions. The \brem losses are given by
\beq
S_1 = -3.8\times 10^{-27}\sqrt{T},
\eeq
the Compton heating and cooling is given by
\beq
S_2 = 4.1\times 10^{-35} (\tx -T)\,\xi ,
\label{eqS2}
\eeq
where $\tx$ is the Compton temperature, and
finally the sum of photoionization heating, line and recombination
continuum cooling is
\beq
S_3 = 10^{-23}{a + b\, (\xi/\xi_0)^c\over 1 + (\xi/\xi_0)^c},
\label{eqS3}
\eeq
where
\beq
a=-{18\over  e^{25 (\log T -4.35)^2}} 
    -{80\over  e^{5.5(\log T -5.2)^2}}
    -{17\over  e^{3.6(\log T -6.5)^2}},
\eeq
\beq
b=1.7\times 10^4\;T^{-0.7}, 
\eeq
\beq
c=1.1-{1.1\over  e^{T/1.8\,10^5}}+{4\times 10^{15}\over T^4}, 
\eeq 
and 
\begin{eqnarray}
\xi_0 &=& {1\over 1.5\; T^{-0.5}+1.5\times 10^{12}\; T^{-2.5}}+\nonumber\\
      &&  {4\times 10^{10}\over T^2}
          \left[1 + {80\over e^{(T-10^4)/1.5\,10^3}}\right].
\end{eqnarray}
Equations~(\ref{eqS2})-(\ref{eqS3}) depend on the ionization parameter
\beq
\xi\equiv {\lbhefphot (r)\over n(r) r^2},
\eeq
where $\lbhefphot (r)$ is the effective accretion luminosity at $r$,
which is evaluated by numerically solving in each shell the balance
equation
\beq
{d\lbhefphot (r)\over dr}=-4\pi r^2 H,
\eeq
with central boundary condition $\lbhefphot (r=0)=\lbh(t)$ given by
equation~(\ref{lbhbol}).  The photoionization+Compton opacity
associated with radiation absorption is then obtained
\beq
\kpho=-{1\over\rho\lbhefphot (r)}{d\lbhefphot (r)\over dr}=
       {4\pi r^2 H(r)\over\rho(r)\lbhefphot (r)}.
\eeq
Finally, the bolometric ISM luminosity is obtained from equation~(\ref{dotE})
as 
\beq
\lrad(r)=4\pi\int_0^r Cr^2 dr.
\label{lrad}
\eeq

The essential physics of this Section is well known. When the
parameter $\xi$ is large thermodynamics guarantees that the gas
temperature approaches the photoionization temperature $\sim 10^{7.3}$
K, but for lower values of $\xi$ the temperature approaches $\sim
10^4$ K, near the peak of the cooling curve.

\subsection{Radiation pressure}

An important ingredient in the modelization of the gas flow evolution
is the radiation pressure due to the accretion luminosity and to the
light produced by the new stars.  In its evaluation the explicit
dependence on time is omitted, since the light travel time in the
regions of interest is small compared to the time-scale on which the
radiative input changes.  Radiation pressure due to {\it electron
scattering} (where neither the photon numbers, nor their energy
change) is computed as
\beq
(\nabla\prad)_{\rm es}=-{\kes\rho\over c}
                          {\lbh+\luv(r)+\lopt(r)+\lrad (r)\over 4\pi r^2},
\label{prad_es}
\eeq
where $\kes=0.35$ in c.g.s. units, and from
equations~(\ref{dEopt})-(\ref{dEUV})
\beq
\luv(r)=4\pi\int_0^r \dEUV r^2 dr,\quad
\lopt(r)=4\pi\int_0^r \dEopt r^2 dr.
\eeq
Note that all the luminosities used in equation~(\ref{prad_es}) are
unabsorbed.

The radiation pressure contribution due to {\it dust opacity} is given
by
\beq
(\nabla\prad)_{\rm dust}=-{\kUV\rho\over c}
                          {\lbhefUV(r)  +\luveff(r)\over 4\pi r^2}
                         -{\kopt\rho\over c}
                          {\lbhefopt (r)+\lopteff(r)\over 4\pi r^2}
                         -{\kIR\rho\over c}{\lir (r)\over 4\pi r^2},
\label{prad_dust}
\eeq
where 
\beq
\lir(r)\equiv L_{\rm BH,UV}^{\rm abs}(r)+L_{\rm BH,opt}^{\rm abs}(r)+
           \luv^{\rm abs}(r) +\lopt^{\rm abs}(r), 
\label{eq_lir}
\eeq
is the infrared luminosity due to recycling of photons absorbed from
the ISM, and we adopt as estimates for (cgs) opacity in three bands
\beq
\kopt={300\over 1 + T/10^4},\quad\kUV= 4\kopt,\quad \kIR={\kopt\over 150},
\eeq
where the temperature dependent denominator is designed to mimic the
destruction of dust at high temperatures (T. Thompson \& B. Draine,
private communication); the dust opacity we are using is likely a
lower bound to the true value, while a more accurate treatment can be
found in Thompson, Quataert \& Murray (2005), and will be implemented
in future explorations. At variance with electron scattering the {\it
effective} luminosities appearing in
equations~(\ref{prad_dust})-(\ref{eq_lir}) take into account
absorption, and are obtained by numerically solving the two lowest
spherically symmetric moment equations of radiative transfer in the
Eddington approximation (e.g., Chandrasekhar 1960):
\beq
{d\luveff\over dr}=4\pi r^2(\dEUV-\kUV\rho\Juveff),\quad
{d\lopteff\over dr}=4\pi r^2(\dEopt-\kopt\rho\Jopteff).
\label{prad_eff1}
\eeq
\beq
{d\Juveff\over dr}=-{3\kUV\rho\luveff\over 4\pi r^2},\quad
{d\Jopteff\over dr}=-{3\kopt\rho\lopteff\over 4\pi r^2},
\label{prad_eff2}
\eeq
The central boundary conditions for stellar luminosities are 
$\luveff(0)=\lduv$, $\lopteff(0)=\ldopt$, $\Juveff (0)=\lduv/16\pi^2R_1^2$ 
and $\Jopteff(0)=\ldopt/16\pi^2R_1^2$ (see Sect. 2.4).
The effective accretion luminosities $\lbhefUV$ and $\lbhefopt$ are
computed with two equations similar to (\ref{prad_eff1}), where the
distributed source term is missing, $J=\lbhef/4\pi r^2$, and in the UV
and optical bands $\lbhefUV (0)=0.2\lbh(t)$ and $\lbhefopt
(0)=0.1\lbh(t)$, respectively.

The last contribution to radiation pressure comes from {\it
photoionization opacity},
\beq
(\nabla\prad)_{\rm photo}=-{\rho\kpho\over c}
                           {\lbhefphot (r)\over 4\pi r^2},
\label{prad_photo}
\eeq
where the photoionization opacity and the absorbed accretion
luminosity are calculated as described in Sect. 2.5.  It is well known
that the radiation pressure on electrons (equation~[\ref{prad_es}])
can significantly affect the gas dynamics when the AGN luminosity
approaches the Eddington limit. Consistently with the findings of
Thompson, Quataert, \& Waxman (2006) we also find that the radiation
pressure on the dust can have a major effect during starburst phases
in retarding the infall of cool gas, thus boosting the mass
transformed into stars and reducing the gas available for accretion
onto the central SMBH.

\subsection{Hydrodynamical equations}

As in CO01, the evolution of the galactic gas flows is obtained
integrating the time--dependent Eulerian equations of hydrodynamics,
where now we have several additional source and sink
terms
\beq
{\pd \rho t}+\nabla \cdot (\rho v)=\alpha\rhos +\drhoII -\drhosp,
\eeq
\beq
{\pd m t}+\nabla\cdot (mv)=-(\gamma-1)\nabla E -\nabla\prad +g\rho -\dot m^+_*,
\eeq
\begin{eqnarray}
{\pd Et}+\nabla \cdot (Ev)&=&-(\gamma-1)\,E\nabla \cdot v + H - C +\\ \nonumber
&&{(\alpha\rhos +\drhoII)(v^2+3\sigast^2)\over 2} +\dEI+\dEII -\dot E^+_*.
\end{eqnarray}
$\rho$, $m$, and $E$ are the gas mass, momentum and internal energy
per unit volume, respectively, and $v$ is the gas velocity.  The ratio
of the specific heats is $\gamma =5/3$, and $g(r)$ is the
gravitational field of the galaxy (stars and dark matter), plus the
contribution of the central SMBH.  The gravitational field is updated
at each time step by considering the SMBH mass growth; for simplicity,
we do not take into account neither the ISM contribution, nor the mass
redistribution due to the stellar mass losses and star formation.  The
total radiative pressure gradient is $\nabla\prad=(\nabla\prad)_{\rm
es}+(\nabla\prad)_{\rm dust}+ (\nabla\prad)_{\rm photo}$ (Sect. 2.6),
while the radiative heating and cooling term $H-C$ is described in
Sect. 2.5.

The energy source term is obtained under the assumption that the
streaming velocity of the source distribution is zero, neglecting the
small contributions of the internal energy of the injected gas, and of
the kinetic energy of stellar wind when compared to the local stellar
velocity dispersion contribution (for the derivation and detailed
discussion of the hydrodynamical equations with moving isotropic or
anisotropic source terms, see Recchi, D'Ercole \& Ciotti 2000).  Note
that the term proportional to the stellar velocity dispersion becomes
dominant near the SMBH, as described in equation~(\ref{jeans}).  The
source terms $\alpha\rhos$ and $\dEI$ of the initial, passively
evolving stellar population are given in
equations~(\ref{alphas})-(\ref{dEI}), while the source terms due to
Type II Supernovae, $\drhoII$ and $\dEII$, are described in Sect. 2.3.

From the numerical point of view, the code is the same as in CO97 and
CO01, however the present simulations are much more difficult to run,
because the stellar density distribution, being more concentrated than
the King model, now injects more gas into the central regions, thus
increasing the gas density and decreasing its cooling time. In
addition, the gas at the galaxy center is gravitationally more bound
(due to higher mass concentration of the new models), and
correspondingly more difficult to expell.  Finally, the numerical
grid spacing has been reduced in order to resolve the central regions
of the galaxy.  In particular, we place the first active grid point
$R_1$ within the Compton radius
\beq
\rx={2G\mbh \mu\mpr\over 3\kb\tx}\simeq 3.6\mu {\mbh\over 10^8\Msun}
    {10^7 {\rm K}\over\tx}\quad {\rm pc},
\label{rx}
\eeq
so that at $R_1$ we can impose the physical condition of a vanishing
{\it thermodynamical} pressure gradient, leading to gas free-fall on
the circumnuclear disk when the radiation pressure is negligible; in
this paper we adopt $\tx= 2.5\times 10^7$ K.  The appropriate values
for radiation pressure at $R_1$ are obtained from the disk treatment
described in Sects. 2.4 and 2.6. Note that in CO01 we were not able
to perform a full simulation with such high resolution, as a
consequence of the higher $\tx$ adopted.

The simulations are realized with a spatially second-order Eulerian
scheme which adopts two staggered grids (for scalar and vector
quantities, see CDPR and CO01 for details), each of them consisting of
120 logarithmically spaced grid points, covering the range 2.5 pc -
200 kpc.  The equations are integrated with a time-splitting scheme,
while the heating and cooling terms in the energy equation are
integrated by using a predictor-corrector scheme, so that the
integration is second order in time. At each simulation time, the
time-step is determined as a fraction of the minimum among the Courant
condition over the grid, and of the others characteristic times
associated with the described physical processes: during the accretion
phases (and subsequents bursts of radiation), it is not infrequent to
have time-steps of the order of 1 yr or less. However, it is important
to note the accretion events are characterized by the intrinsic
time-scale related to equation~(\ref{rx}) by
\beq
t_{\rm X}\equiv {\rx\over c_{\rm X}}\simeq 1.22\,10^4 \mu^{3/2} 
                {\mbh\over 10^8\Msun}
                \left({10^7 {\rm K}\over\tx}\right)^{3/2}\quad {\rm yr},
\label{timex}
\eeq
where $c_{\rm X}$ is the isothermal sound velocity associated with the 
Compton temperature.

\section{Model evolution}

We now show the main properties of a representative model
characterized by an initial stellar mass $\mast= 4.6\times
10^{11}\Msun$, a FP effective radius $\re=6.9$ kpc and aperture
velocity dispersion $\sigma_{\rm a}=260$ km s$^{-1}$ (leading to an
expanded initial condition of $\re = 10.35$ kpc and $\sigma_{\rm
a}=235$ km s$^{-1}$), total dark-to-visible mass ratio $\RM =5$ and
dark-to-visible scale-length ratio $\beta =5.22$ (corresponding to an
identical amount of stellar and dark matter within the half-light
radius).  The initial SMBH mass follows the present day Magorrian
relation, i.e., $\mbh\simeq 10^{-3}\mast$. We remark again that this
model galaxy is not fully appropriate as an initial condition for a
cosmological simulation, because its parameters are fixed to reproduce
an early-type galaxy similar to those observed in the local universe,
and also because we set outflow boundary conditions at the galaxy
outskirsts ($\sim 200$ kpc): from this point of view, the simulations
represent an isolated elliptical galaxy (note for example that in the
present context we are not considering the effects of possible merging
on the galaxy evolution).  We adopted this procedure to adhere to the
standard approach followed in "cooling-flow" simulations, while in
future explorations we will address in a more consistent way the
problem of the galaxy structural and dynamical modifications due to
star formation and mass redistribution over an Hubble time, and the
compatibility of the obtained galaxies with the present-day scaling
laws of elliptical galaxies.  We remark that the model presented
is just one out several tens of runs that have been made,
characterized by different choices of the parameters (often outside
the currently accepted ranges), for example with high/null DM,
enhanced/suppressed star formation, vanishing efficiencies, and so
on. The obtained model evolution can be very different, some of them
leading to galaxies in a permanent wind state, or galaxies with
extremely massive final SMBHs. A discussion of these issues is
deferred to successive works.

The initial conditions are represented by a very low density gas at
the local virial temperature.  The establishment of such
high-temperature gas phase at early cosmological times is believed to
be due to a "phase-transition" when, as a consequence of star
formation, the gas-to-stars mass ratio was of the order of 10\% and
the combined effect of SNIa explosions and AGN feedback became
effective in heating the gas and driving galactic winds. Several
theoretical arguments and much empirical evidence, such as galaxy
evolutionary models and the metal content of the Intra Cluster Medium
(ICM) support this scenario (e.g., Renzini et al. 1993; OC05; Di
Matteo et al. 2005). For the reasons above, in the simulation here
presented (as well as in all others simulations not shown), we assume
that the age of the galaxy stellar component at the beginning of the
simulation is 2 Gyr old, and the simulations span 12 Gyr, so that the
cosmic time at the end of the simulations is 14 Gyr.

\subsection{Luminosities}

\begin{figure}
\includegraphics[angle=0.,scale=1]{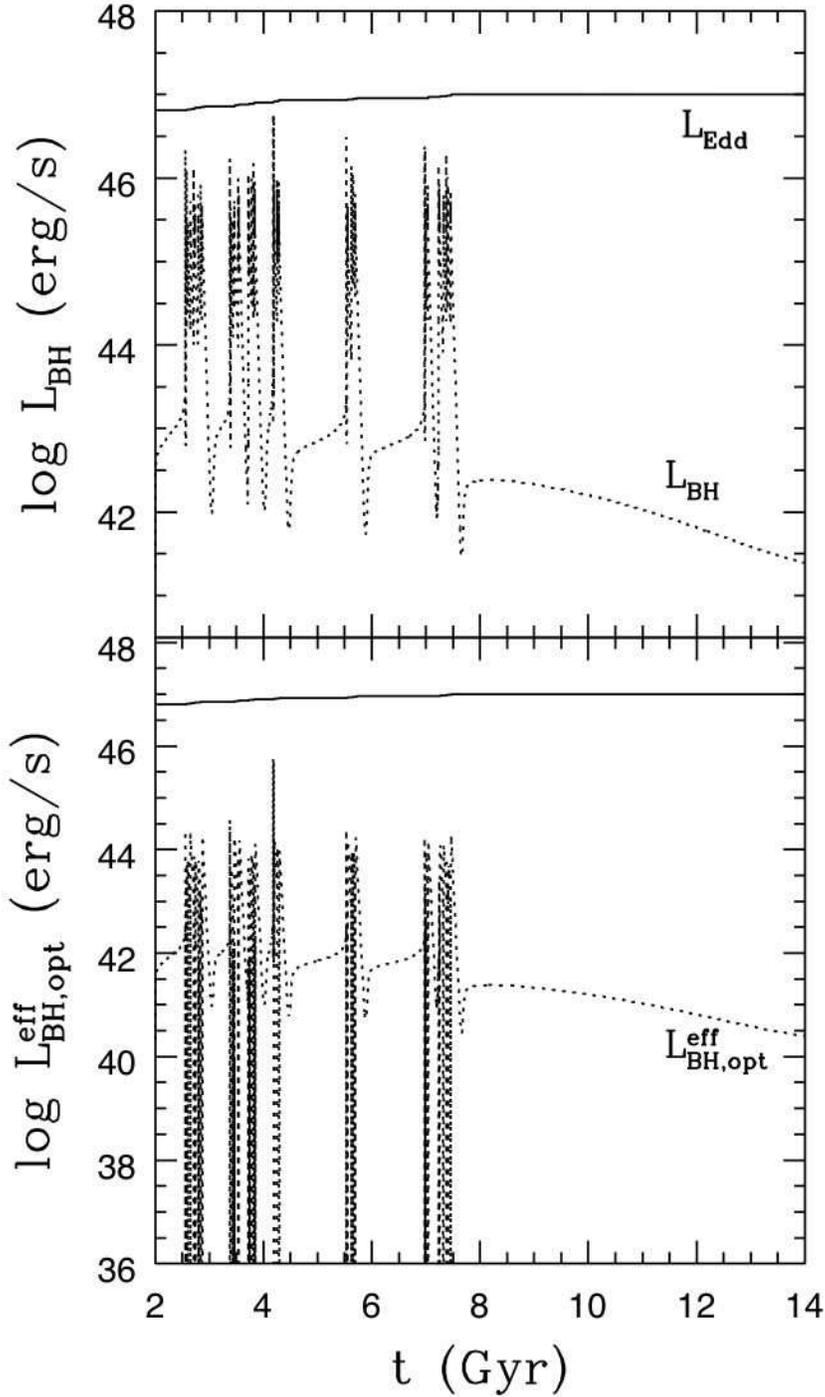}
\caption{Dotted lines are the bolometric accretion luminosity (top
panel) and the optical SMBH luminosity corrected for absorption
$\lbhefopt$, i.e, as would be observed from infinity (bottom panel,
see Sect. 2.6); we recall that at the center we fixed $\lbhefopt
(R_1)=0.1\lbh$. The almost horizontal solid line is $\ledd$.}
\label{lbh}
\end{figure}
\begin{figure}
\includegraphics[angle=0.,scale=0.9]{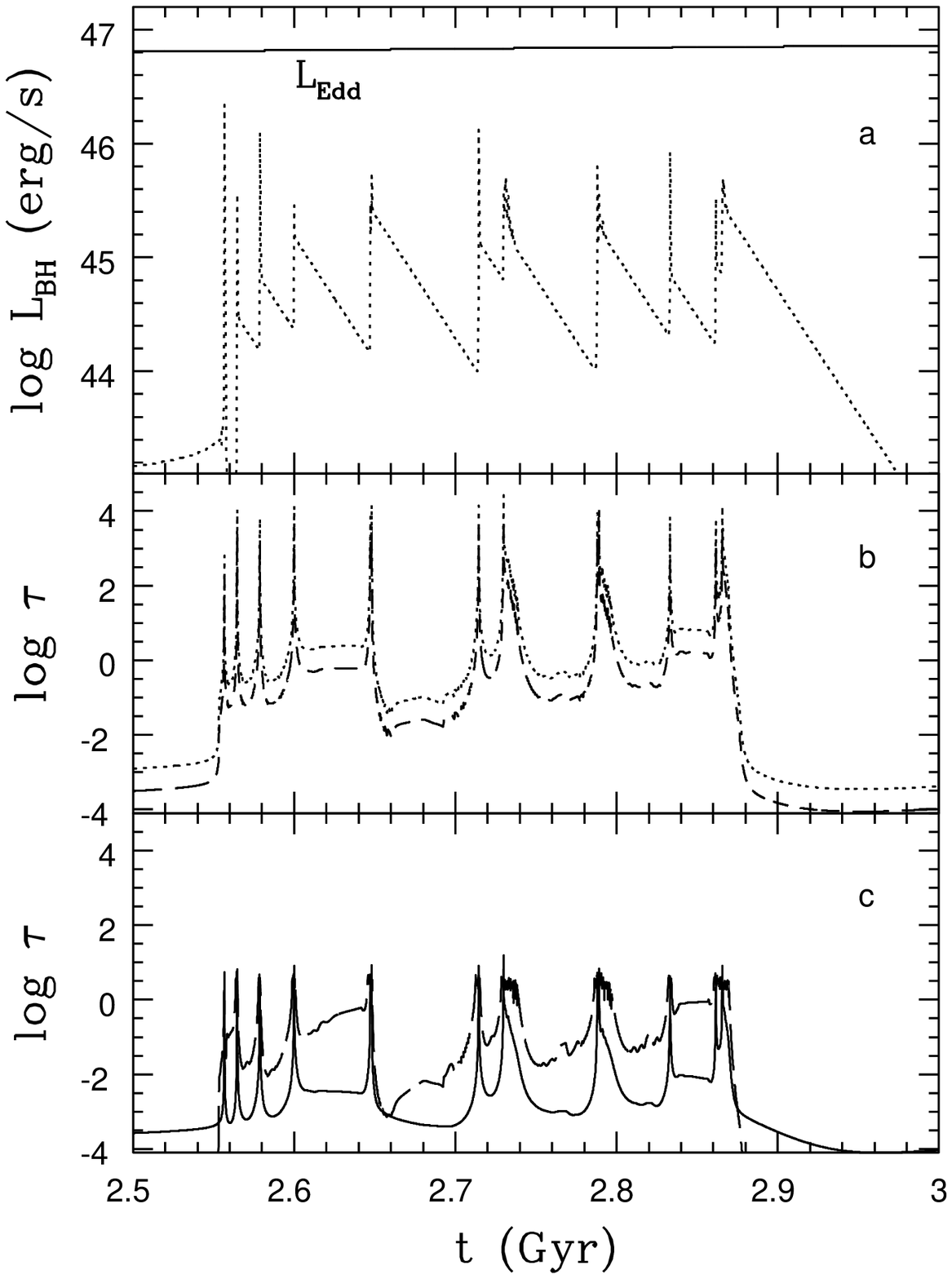}
\caption{Panel a: time expansion of 500 Myrs of top panel in
Fig.~\ref{lbh}, showing the first major burst.  The dotted line is the
{\it bolometric} accretion luminosity.  The widths of the spikes is
typically of the order of $\sim 0.1-1$ Myr.  Total opacities (defined
as $\tau= \int_0^{R_t}\kappa\rho dr$, where $R_t=200$ kpc) 
of dust on UV (dotted) and
optical (dashed) luminosities (panel b), and electron scattering
(solid), and photoionization opacity (long dashed line) (panel c).
Dust opacity on the IR recycled radiation would be a line parallel to
those in panel b, but $\sim$ two order of magnitude lower (see
equation [52]).  Note that the vertical scale is the same in panels
$b$ and $c$.}
\label{lbhRES}
\end{figure}

A first, important result of the new models is that overall the main
properties of the CO01 models are confirmed. After a first
evolutionary phase in which a galactic wind is sustained by the
combined heating of SNIa and thermalization of stellar velocity
dispersion, the central "cooling catastrophe" commences. In absence of
the central SMBH a "mini-inflow" would be then established, with the
flow stagnation radius (i.e., the radius at which the flow velocity is
zero) of the order of a few hundred pc to a few kpc.  These
"decoupled" flows are a specific feature of cuspy galaxy models with
moderate SNIa heating (Pellegrini \& Ciotti 1998). However, after the
central cooling catastrophe, the feedback caused by photoionization
and Compton heating strongly affects the subsequent evolution, as can
be seen in Fig.~\ref{lbh} where we show the luminosity evolution of
the central AGN with time-sampling of $10^5$ yrs.  The bolometric
luminosity (top panel) ranges between roughly 0.1 to 0.001 of the
Eddington limit (the almost horizontal solid line) at peaks in the
SMBH output but, since obscuration is often significant, the optical
accretion luminosity as seen from infinity can be much lower still
(bottom panel).  But the central quasar is not always obscured and we
see, in the lower panel of Fig.~\ref{lbh}, that the optical luminosity
reaches $\sim 10^{44}$ erg s$^{-1}$ in numerous bursts.  As already
found in CO01, the major AGN outbursts are separated by increasing
intervals of time (set by the cooling time), and present a
characteristic temporal substructure, whose origin is due to the
cooperating effect of direct and reflected shock waves.  At $t\simeq
8$ Gyrs the SNIa heating, also sustained by a last strong AGN burst,
becomes dominant, a global galactic wind takes place and the nuclear
accretion switches to the optically thin regime. Note that a further
reduction of the accretion luminosity during this phase would be
otained if considering ADAF accretion instead of standard accretion.

The temporal substructure of the first major burst is revealed in
Fig.~\ref{lbhRES}a, where we show a blow-up of 500 Myr of top panel
Fig.~\ref{lbh}, starting at 2.5 Gyr: the time extent of each of the
sub-bursts (for example, when $\lbh > 10^{45}$ erg s$^{-1}$) is of the
order of $\lsim 1$ Myr. In Fig.~\ref{lbhRES}b,c we also show the time
evolution of the different model opacities during the same burst,
where the transition from optically thin to the optically thick and
back to thin phase is apparent.

\begin{figure}
\includegraphics[angle=0.,scale=1]{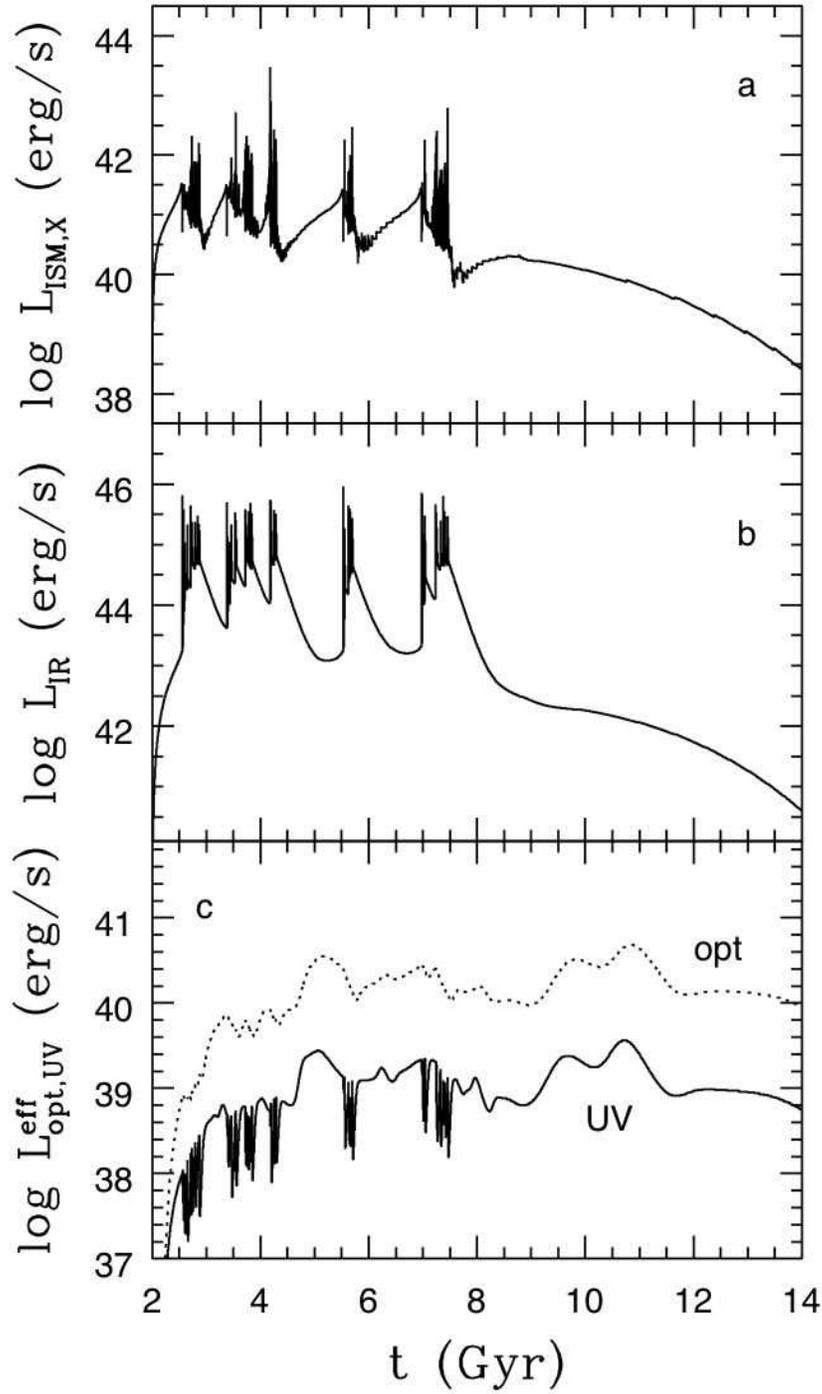}
\caption{The galaxy X-ray coronal luminosity $\lx$ (top), 
recycled infrared $\lir$ (middle), and the starburst UV and optical
luminosities (bottom), corrected for absorption.}
\label{lum}
\end{figure}

In Fig.~\ref{lum}a we show the coronal X-ray luminosity $\lx$ (emitted
by gas at $T\geq 5\times 10^6$ K), due to the hot galactic atmosphere
integrated within $10\re$.  In particular, $\lx$ falls in the range
commonly observed in massive early-type galaxies, with mean values
lower than the expected luminosity for a standard cooling-flow model.
In Fig.~\ref{lum}b we show instead the estimated IR luminosity $\lir$
due to the reprocessing of the radiation emitted by the new stars and
by the SMBH and absorbed by the ISM inside $10\re$ (equation [51]).
The simulations show that the bulk of the reprocessed radiation comes
from AGN obscuration, while the lower envelope is set by radiative
reprocessing from the new stars.  The very high luminosity peaks
($\lir\sim 10^{45-46}$ erg s$^{-1}$) correspond to one component of
the SCUBA sources seen at $z\sim 2$ (e.g., see Pope et al. 2006).
Finally, in Fig.~\ref{lum}c we show the temporal evolution of the
optical and UV luminosities of the starbursts corrected for
absorption.  Overall, Figs.~\ref{lum}bc show that a large fraction of
the starburst luminosity output occurs during phases when shrouding by
dust is significant (e.g., see Rodighiero et al. 2007), i.e. the model
would be observed as an IR source with UV and optical in the range
seen in brighter E+A sources.

\begin{figure}
\includegraphics[angle=0.,scale=1]{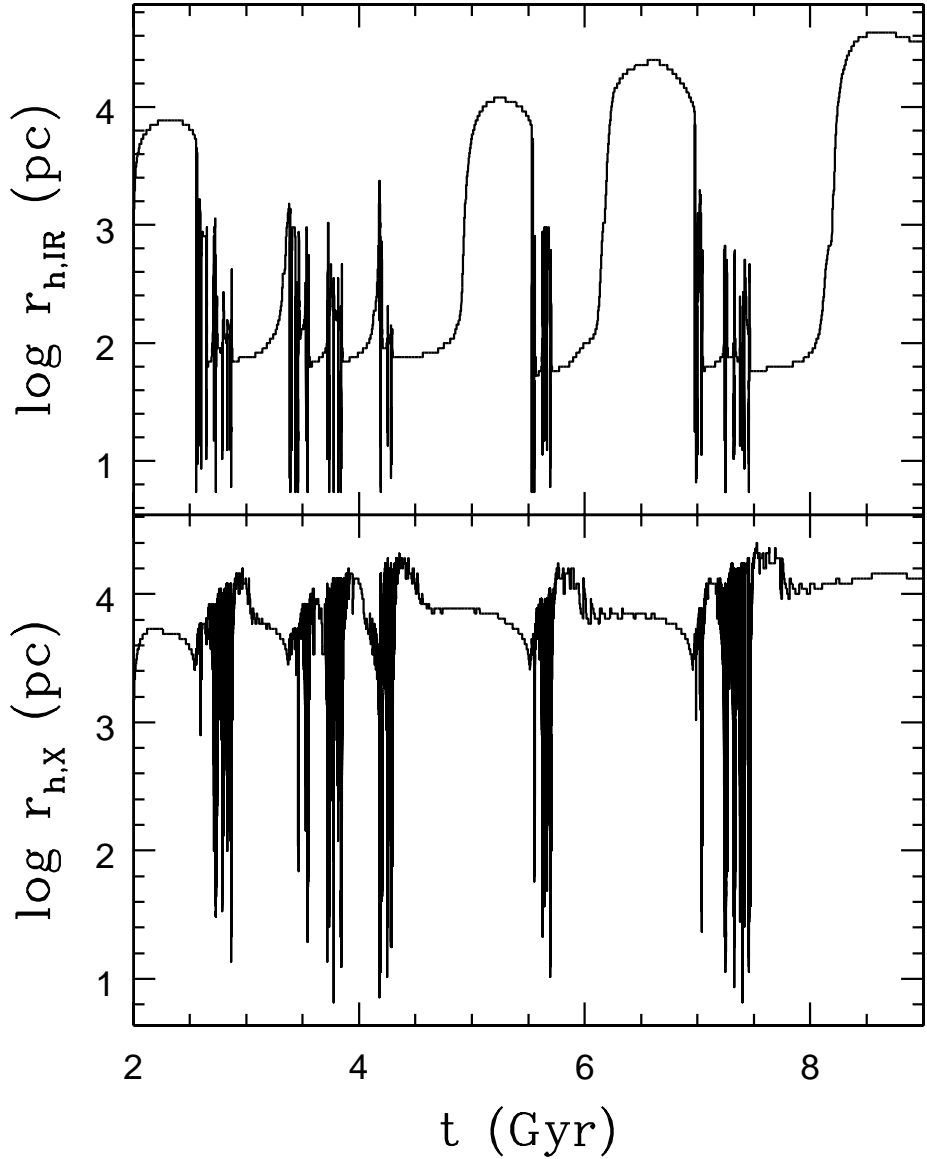}
\caption{Time evolution of the (volume) half-light radius of the X-ray
ISM luminosity (bottom panel) and IR luminosity (top panel) during the
bursting phases.  Small radii correspond to the high-luminosity peaks,
and the predicted SCUBA-like sources should be of linear size $\sim
10^{2.5}$ pc.}
\label{rhalf}
\end{figure}

The spatial radius within which half of the IR and X-ray luminosities
are emitted changes dramatically with time, and as a function of the
total emitted luminosity.  This is apparent from Fig.~\ref{rhalf},
where we show the time evolution of $r_{\rm h,X}$ and $r_{\rm
h,IR}$. Note the large variation of the size of the emission regions:
in particular, the smallest values of the half-light radii correspond
to the peaks of the associated luminosities. During strong radiation
outbursts, the emitting regions are so small ($\sim 10$ pc or less)
that virtually the entire X-ray and IR luminosities would be seen as
emitted by the central source; these phases also correspond to phases
of high obscuration, with optical depth of the order of 100 (e.g.,
Imanishi et al. 2007). Instead, during the late-time hot accretion
phase $r_{\rm h,X}$ reaches values commonly observed in elliptical
galaxies, and $R_{\rm h,IR}$ is even larger. When the IR luminosity is
as large as seen in the SCUBA observations (Pope et al. 2006), we
predict that the characteristic sizes will be of the order of $\sim
10^{2.5}$ pc.
\begin{figure}
\includegraphics[angle=0.,scale=0.9]{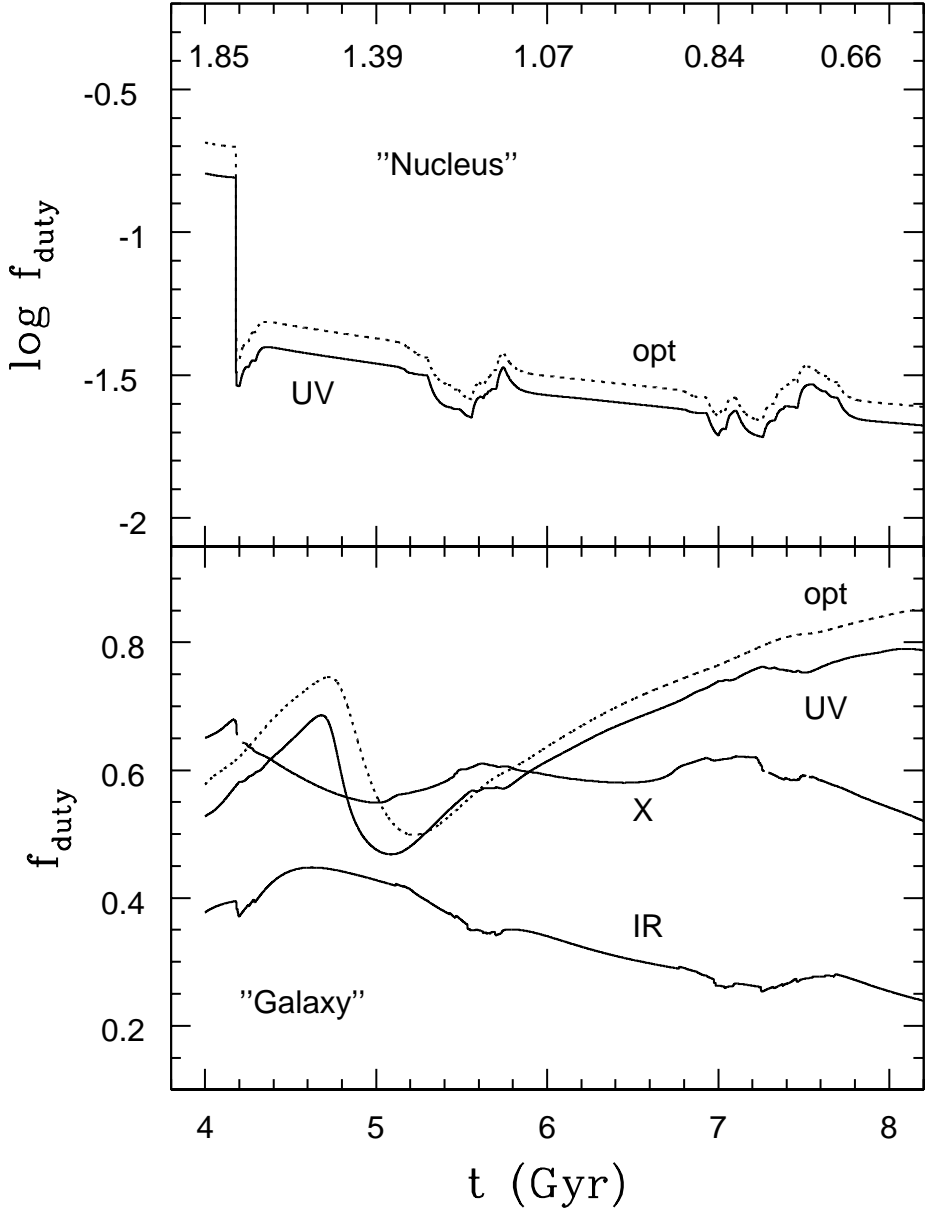}
\caption{Time evolution of duty cycles as computed from equation~(61),
with a time-window $\Delta t = t/2$. Top panel: duty-cycle of
$\lbhefUV$ (solid) and $\lbhefopt$ (dotted); the top axis shows the
corresponding redshift (e.g., Spergel et al. 2006).  We see that these
systems would be observed from afar in the (rest-frame) optical or UV
as quasars several percent of the time.  Bottom panel: duty-cycle of
the starburst $\luveff$ (solid), $\lopteff$ (dotted), of the ISM X-ray
luminosity (computed in a volume excluding the inner 100 pc), and of
the recycled IR luminosity $\lir$.}
\label{duty}
\end{figure}

An important quantity associated with the time evolution of the
various luminosities is their {\it duty cycle}. As in CO01 for a given
luminosity $L(t)$, we define the associated duty-cycle\footnote{
Defined in this way $f_{duty}$ would be the fraction of the time a
system would be in a high-luminosity state, $L_h$, and $1-f_{duty}$
would be the fraction of the time it spent in a low-luminosity state
$L_l$, if it were to oscillate between these two states and $L_l/L_h
<< 1$.} over a period of time $\Delta t$
\beq
f_{duty}\equiv
{[\int^t_{t-\Delta t} L(t')dt']^2\over \Delta t\int^t_{t-\Delta t} L^2(t')dt'}.
\eeq
In Fig.~\ref{duty} we show the resulting values for a time dependent
temporal window $\Delta t = t/2$. In the top panel we show the
(logarithmic) value for the effective optical and UV AGN luminosities
(with negative peaks corresponding to values as low as 0.01 or less),
while during the smooth accretion of the last Gyrs the values flatten
to unity. In the bottom panel the duty-cycles correspond to the
starburst optical and UV luminosities, and show a larger and less
fluctuating values $\gsim 0.5$, in agreement with observational
results (e.g., see Cimatti et al. 2002).  In the same panel we also
show the duty-cycle of the X-ray ISM luminosity computed outside a
sphere of radius 100 pc, so as to exclude the ISM luminosity
fluctuations produced by direct AGN heating on the surrouding ISM. In
other words, the X-ray duty cycle refers to the bulk of the galaxy
body, and should give an indication of the expected fraction of
significantly disturbed galaxies in coronal X-rays.  Quite obviously,
the derived values depend on the adopted sampling time interval
$\Delta t$. For example, by fixing the sampling time to $\Delta =100$
Myrs, the values in the bottom panel are almost unchanged, while the
AGN duty-cycles becomes (during peaks of activity) as small as $\sim
10^{-3}$.  This is consistent with the temporal substructure of the
major bursts (see Fig.~\ref{lbhRES}) At the opposite end, taking all
the time interval spanned by the simulation, the AGN duty-cycle (both
in UV and optical) is $\sim 10^{-2}$, the IR is $\sim 0.2$, while the
starburst duty-cycle is $\sim 0.8$ and that of the global ISM X-ray
$\sim 0.4$. We stress here that the duty-cycles are computed by the
code calculating the luminosity values at {\it each} time-step, that
is usually of the order of the year or even less.
\begin{table}
\centering
\caption{Column 1 gives the adopted criterion to quantify fraction of
the time spent in the high-luminosity states of accretion, in terms of
the emitted SMBH bolometric luminosity and the current Eddington
luminosity. In Column 2 we give the time percentage (calculated over
the model bursting period) $F_{\rm ON}\equiv \Delta t_{\rm ON}/\Delta
t_{\rm bursting}$, with $\Delta t_{\rm bursting}=5.5$ Gyr (see
Fig.~\ref{lbh}). Column 3 gives the time percentage of obscuration
(more than 2 mag in the rest-frame UV) calulated from the ratio
$\lbhefUV/0.2\lbh$ over the time interval $\Delta t_{\rm
ON}$. Finally, $F_{\rm QSO}\equiv F_{\rm ON}\times$ (1-Column 3) gives
the fraction of the bursting period during which there is less than 2
mag of obscuration in the rest frame UV or less than 1.2 mag of
obscuration in the rest-frame optical. This corresponds roughly to
X-ray obscuration of $\approx 10^{22}$ atoms cm$^{-2}$.}
\begin{tabular}{cccc}
\\
\hline
$\lbh/\ledd$  & $F_{\rm ON}$ & $\Delta t_{\rm obsc}/\Delta t_{\rm ON}$ & 
$F_{\rm QSO}$\\
\hline
$>0.100$ & $0.09\%$ & $80\%$ & $0.018\%$\\
$>0.030$ & $1.51\%$ & $94\%$ & $0.096\%$\\
$>0.010$ & $9.78\%$ & $48\%$ & $5.1\%$\\
$>0.003$ & $22.48\%$ & $36\%$ & $14.4\%$\\
$>0.001$ & $28.61\%$ & $32\%$ & $19.5\%$\\
\hline
\end{tabular}
\end{table}

Duty cycles can be also defined in a different way. For example, in
Table 1 we focus in particular on the estimated time fraction spent by
the central SMBH shining at a given fraction of Eddington luminosity,
also considering the obscured accretion phases. Overall, in the
bursting phase ($1\lsim z\lsim 3$), the duty-cycle of the SMBH in its
"on" phase is of order precents and it is unobscured approximately
one-third of the time. We found it interesting that these figures,
obtained from hydrodynamical simulations, can be positively compared
with observations (e.g., see Gilli, Comastri \& Hasinger 2007;
Martinez-Sansigre \& Rawlings 2007).

\subsection{Mass budgets}

\begin{figure}
\includegraphics[angle=0,scale=0.8]{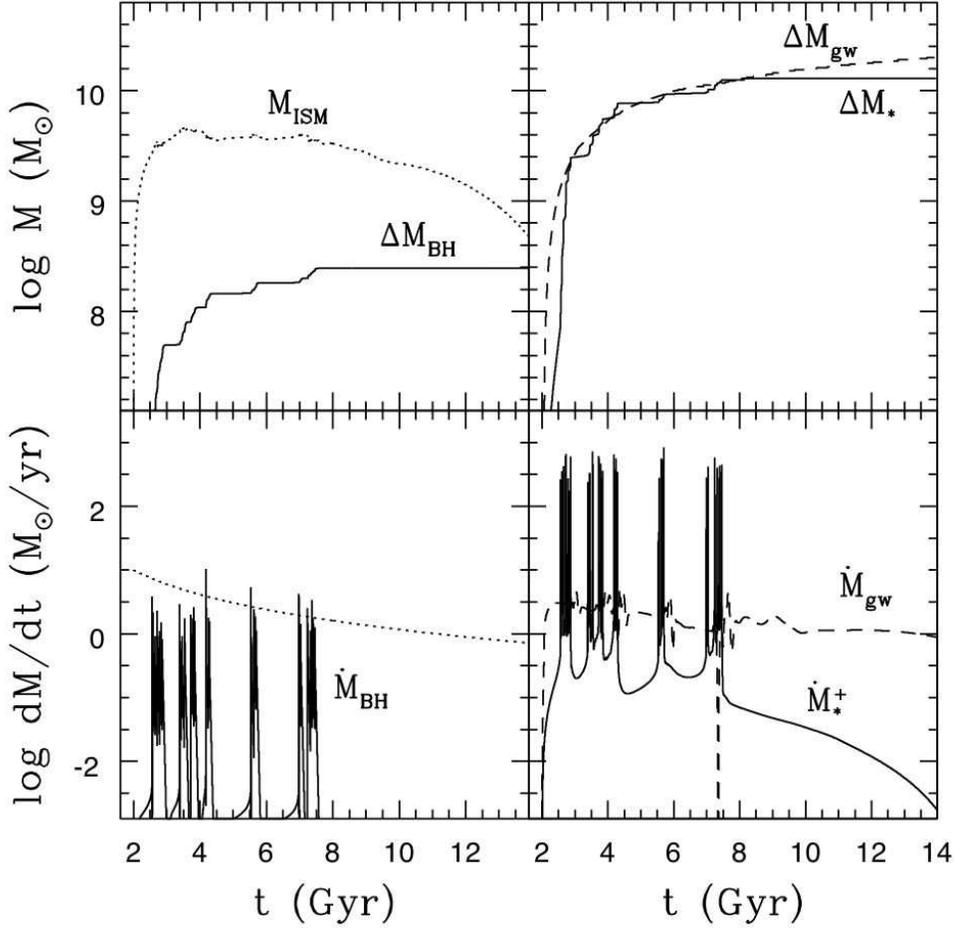}
\caption{Mass budget evolution.  Top left panel: total hot gas mass in
the galaxy (within $10\re$, $M_{\rm ISM}$, dotted line), and accreted
mass on the central SMBH ($\Delta\mbh$). Top right: mass lost as a
galactic wind at $10\re$ ($\Delta M_{\rm gw}$, dashed line), and total
mass of new stars ($\Delta M_*$) formed according to
equation~(\ref{drhosp}). Bottom left panel: mass return rate from the
evolving stellar population (as given by volume integral of
equation~[\ref{alphas}], dotted line), and mass accretion rate on the
central SMBH ($\dot\mbh$, equation~[26]). Bottom right: galactic wind
mass loss rate at $10\re$ ($\dot M_{\rm gw}$, dashed line), and
instantaneous, volume integrated, star formation rate. Note that for
$t >8$ Gyrs the mass lost as a galactic wind is almost coincident with
the mass input from evolving stars.}
\label{mass}
\end{figure}

In Fig.~\ref{mass} we show the time evolution of some of the relevant
mass budgets of the model, both as time-integrated properties and
instantaneous rates.  At the end of the simulation the total ISM mass
in the galaxy is $\sim 5\times 10^8\Msun$, while the SMBH mass
increased by $2.5\times 10^8\Msun$, thus reaching a final mass of
$\sim 7\times 10^8\Msun$: a model with a smaller initial SMBH mass
would accrete less, thus maintaining the Magorrian relation even
better.  The SMBH mass accretion rate strongly oscillates as a
consequence of radiative feedback, with peaks of the order of 10 or
(more) $\Msun$/yr, while during the final, hot-accretion phase the
almost stationary accretion is $\lsim 10^{-4}\Msun$/yr: this value is
close to the estimates obtained for the nuclei of nearby galaxies
(Pellegrini 2005).  Note that in the last 6 Gyrs the SMBH virtually
stops its growth, while the ISM mass first increased due to the high
mass return rate of the evolving stellar population, and then
decreases due to the global galactic wind induced by SNIa.  During the
entire model evolution, $\gsim 10^{10.5}\Msun$ of recycled gas has
been added to the ISM from stellar mass losses.  Approximately
$2.1\times 10^{10}\Msun$ has been expelled as a galactic wind, while
$\sim 1.4\times 10^{10}\Msun$ has been transformed into new stars, so
that only 0.7\% of the recycled gas has been accreted onto the central
SMBH, and the central paradox of the mass budget is automatically
resolved. 

In the present simulation, approximately twice of the total mass
accreted onto the central SMBH is expelled as a disk wind, while the
final mass of the disk, in stellar remnants, sums up to $\mrem\sim
2.8\times 10^5\Msun$.  It is important to stress that an identical
model without SMBH feedback (i.e., $\eps=0$ in equation [33]), but
with the same star formation treatment of the model described in this
paper, produced a SMBH of final mass $\gsim 10^{10}\Msun$, while the
total mass in new stars was reduced to $\sim 3\times 10^9\Msun$.  In
addition, this "ad hoc" model does not present fluctuations in the
starburst and ISM X-ray luminosities, thus showing the vital
importance of SMBH feedback in the overall results.
Tests of numerical convergence were performed to determine 
the extent to which the results quoted in this paper would be altered as 
one  increased the spatial and temporal resolution. To this end we 
reduced the grid spacing by a factor of 1.5 and then a factor of 2.0. 
Almost all results changed at the level of a few percent or less 
(with the numerical uncertainty thus far below the level of the 
uncertainty in the physical modeling). The largest change found in the 
highest resolution run was in the growth of the  central SMBH mass with 
this growth being reduced by $\sim 14$\% .  This is in the direction 
to make the final results conform more closely to the Magorrian relation 
with regard to $\Delta\mbh/\Delta\mast$ (see Fig.~[\ref{mass}]).

\begin{figure}
\includegraphics[angle=0,scale=0.95]{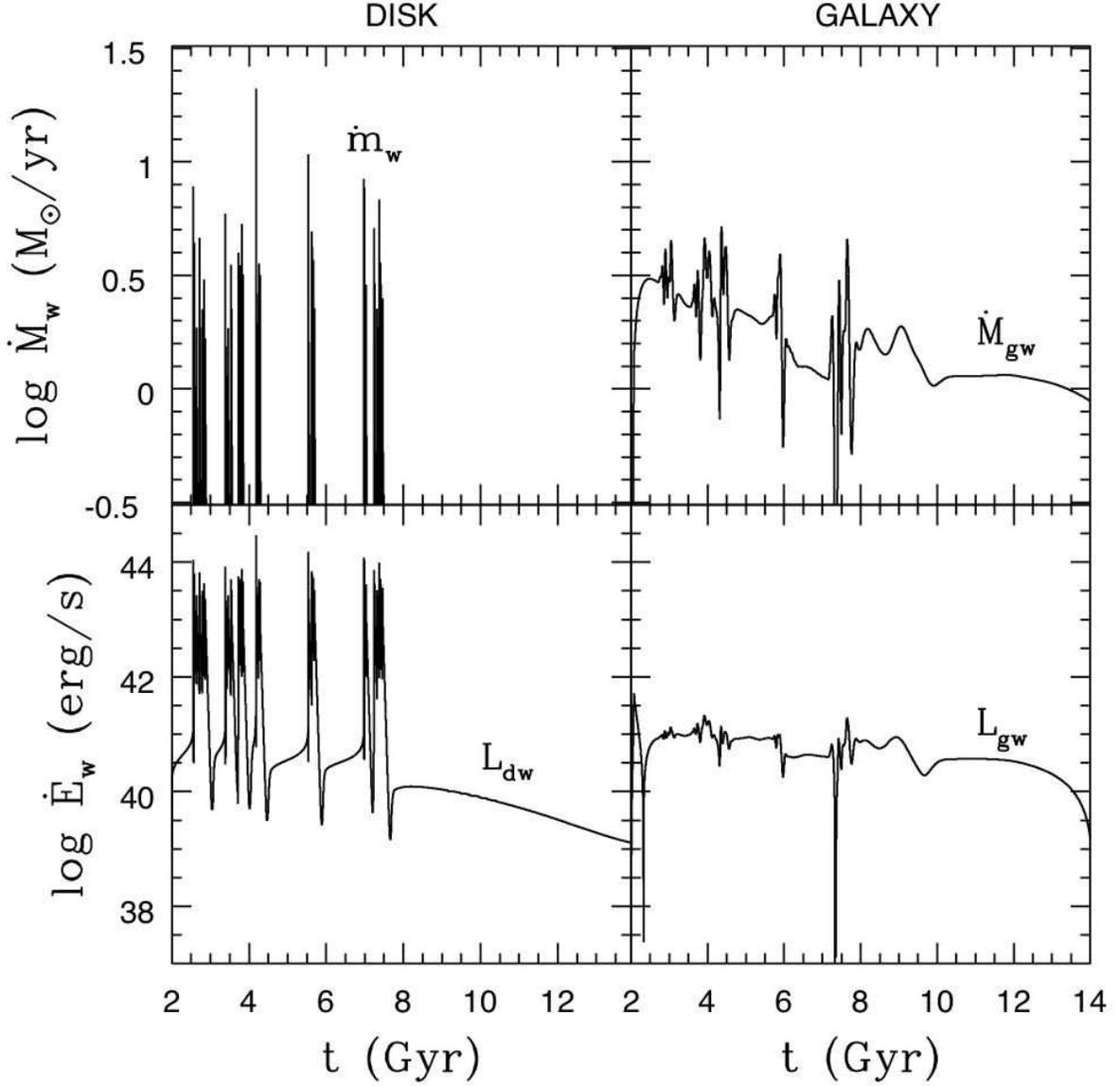}
\caption{Top panels: temporal evolution of the disk wind mass rate
(equation~[\ref{eqmdiskw}], left) and the galactic wind mass rate
computed at $10\re$, $\dot M_{\rm gw}\equiv 4\pi
(10\re)^2\rho(10\re)v(10\re)$ (right).  Bottom panel: the
corresponding mechanical luminosities, i.e., the kinetic energy that
would be injected from the disk in the galaxy interstellar medium
($L_{\rm dw}$), and by the galaxy in the ICM, $L_{\rm gw}\equiv \dot
M_{\rm gw}(10\re)v^2(10\re)/2$.}
\label{lwind}
\end{figure}

Two important quantities associated with the mass budget of the model
are the mechanical (non-relativistic) wind luminosity of the disk as
given by equation~(\ref{eqldwin}), and the mechanical luminosity of
the galaxy, i.e. the kinetic energy carried away by the global galactic
wind (Fig.~\ref{lwind}, bottom panels, while the corresponding mass
rates are shown in the top panels).  Time integration of these two
mechanical luminosities over the entire model evolution revealed that
the disk wind would deposit $\sim 2.2\times 10^{59}$ erg in the galaxy
ISM, while the galactic wind would inject in the ICM $\sim 1.3\times
10^{58}$ erg.
\begin{figure}
\includegraphics[angle=0.,scale=0.9]{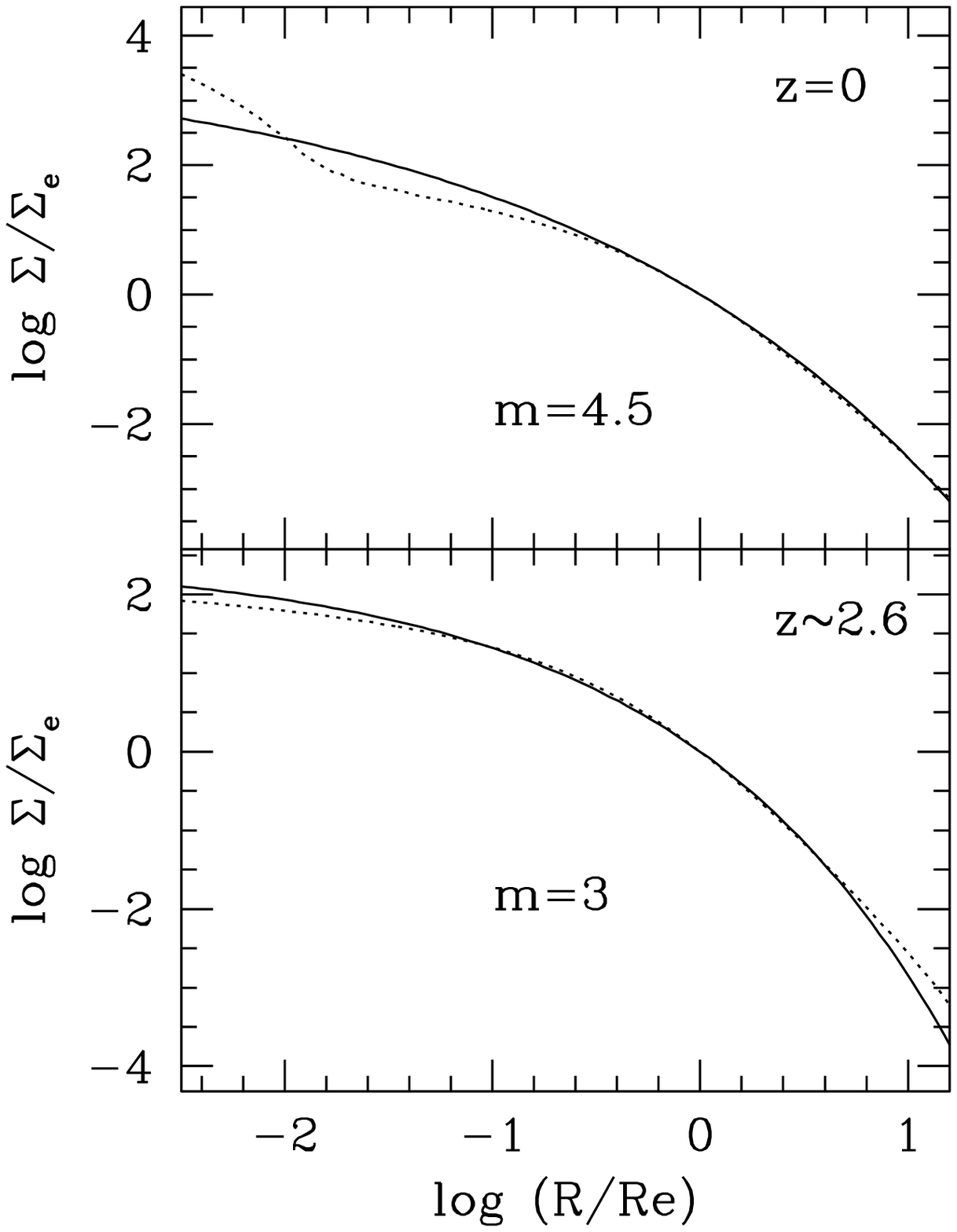}
\caption{Dotted lines are the projected surface density of the model
shortly after the beginning of the simulation ($t=2.5$ Gyr, $z\sim
2.6$, bottom panel), and at $t=13.5$ Gyr ($z=0$, top panel),
normalized to the surface density at the effective radius. The
relation between age and redshift holds for standard Cosmology (e.g.,
Spergel et al. 2006). Solid lines are the best-fit Sersic law. The
effective radius contracted from $\sim 9.2$ kpc to $\sim 8.4$ kpc,
while the surface density $\Sigma_{\rm e}$ increased from $\sim
3\times 10^{22}$ to $\sim 3.6\times 10^{22}$ protons per cm$^{-2}$.}
\label{surf}
\end{figure}
The star formation rate during the periods of feedback dominated
accretion oscillates from $0.1$ up to several hundreds (with peaks
near $10^3$) $\Msun$ yr$^{-1}$, while it drops monotonically from
$10^{-1}$ to $\lsim 10^{-3}$ $\Msun$ yr$^{-1}$ in the last 6 Gyrs of
quiescent accretion (see Fig.6). As already mentioned above, these
violent star formation episodes (with SMBH accretion to star formation mass
ratios $\sim 10^{-2}$ or less) are induced by accretion
feedback\footnote{However, bursting star formation is not necessarily
associated with AGN feedback (e.g., see Kr\"ugel \& Tutukov 1993).},
and are spatially limited to the central $10-100$ pc; thus, the bulk of gas 
flowing to the center is consumed in the starburst. These findings
are nicely supported by recent observations (e.g, see Sect. 5 in Lauer
et al. 2005, see also Davies et al. 2007). Note that the "age" effect
of the new stars on the global stellar population of the galaxy is
small, as the new mass is only 3\% of the original stellar mass, and
it is virtually accumulated during the first Gyrs (see
Fig.~\ref{mass}), so that the mass-weighted age of the final model is
of the order of 12 Gyrs. The half-mass radius of the final stellar
distribution (without considering adiabatic contraction, nor the
reduction of the stellar mass distribution due to galactic winds, see
Section 4) contracts by $\sim 16$\%, just due to the addition of the
new stars in the central regions of the galaxy. This is made apparent
in Fig.~\ref{surf}, where we show the final spatial density profile of
the system, together with its projection and the best-fit obtained
with the Sersic (1968) law
\beq
\Sigma(R)=\Sigma_0 {\rm e}^{-b(m)(R/\re)^{1/m}},
\eeq
where $b(m)=2m-1/3+4/(405m)$ (Ciotti \& Bertin 1999). As expected, the
profiles show an increase of the best-fit Sersic parameter $m$, due to
the mass accumulation in the central regions. Remarkably, the final
value of $m$ is within the range of values commonly observed in
ellipticals: however, in the final model we note the presence of a
central ($\sim 30$ pc) nucleus originated by star formation which
stays above the best fit profile, similar to the light spikes
characterizing "nucleated" galaxies (e.g., see Graham \& Driver 2005,
Lauer et al. 2005).

\subsection{Hydrodynamics}

\begin{figure}
\includegraphics[angle=0.,scale=0.9]{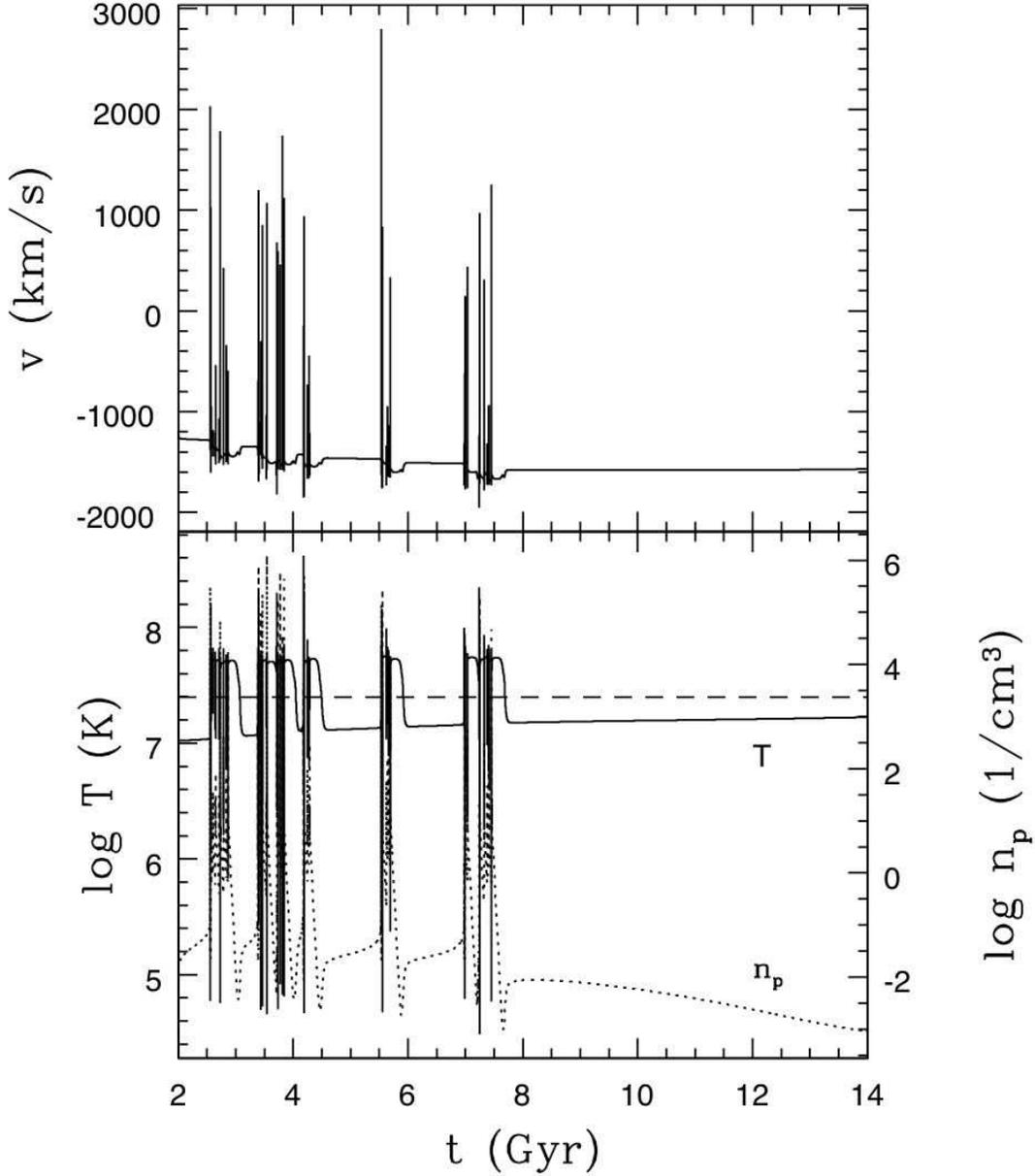}
\caption{Top panel: gas velocity at 5 pc from the SMBH. Note how the SMBH
growth affects the lower envelope of velocity values.  Bottom panel:
Gas number density (dotted line, scale on the right axis) and
temperature at 5 pc from the SMBH (solid line). Low-temperature,
high-density phases end when accretion luminosity $\lbh$ increases
sharply heating the ambient gas to a high-temperature, low-density
state. The horizontal dashed line is the model Compton temperature 
$\tx=2.5\times 10^7$ K.}
\label{cen}
\end{figure}

\begin{figure}
\includegraphics[angle=0.,scale=1]{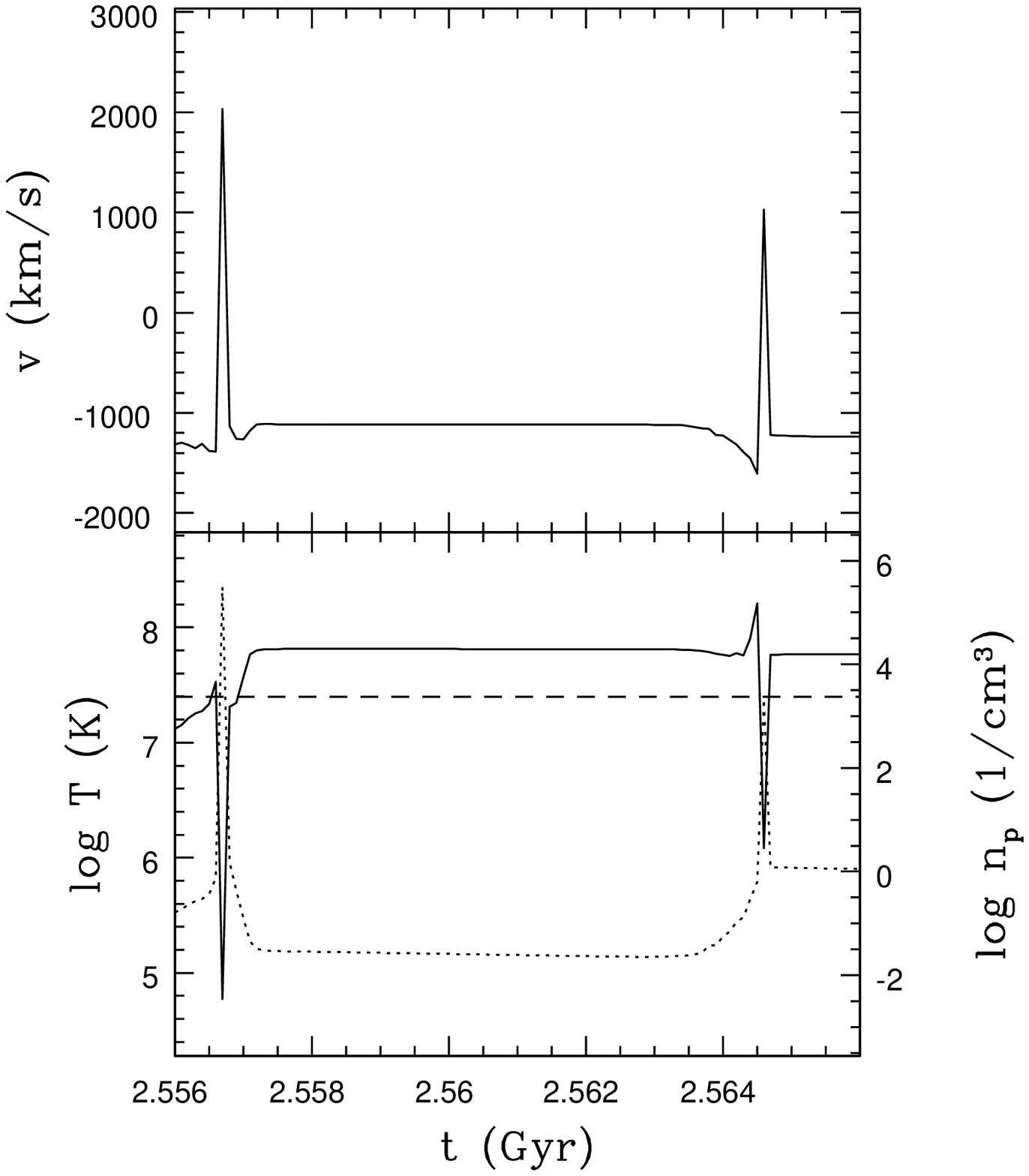}
\caption{Time expansion of the first two feedback events of
the initial major burst in Fig.~\ref{cen} (whose total time extent is 
$\sim 400$ Myr) at time resolution $10^5$ yr; the horizontal 
dashed line is the model Compton temperature. The complementary 
behavior of $\rho$ and $T$ is apparent.}
\label{cenRES}
\end{figure}

In Fig.~\ref{cen} we show the temperature and density in the central
regions of the model: note how the SMBH bursts heat the central gas,
causing the density to drop, and launching gas at positive velocities
of the order of thousands km s$^{-1}$ (this can better appreciated in
Fig.~\ref{cenRES}, where we show a time blow-up of the first two SMBH
feedback events). The Compton temperature is the horizontal dashed
line, and during the bursts the local gas is heated above this limit.


\begin{figure}
\includegraphics[angle=0.,scale=1]{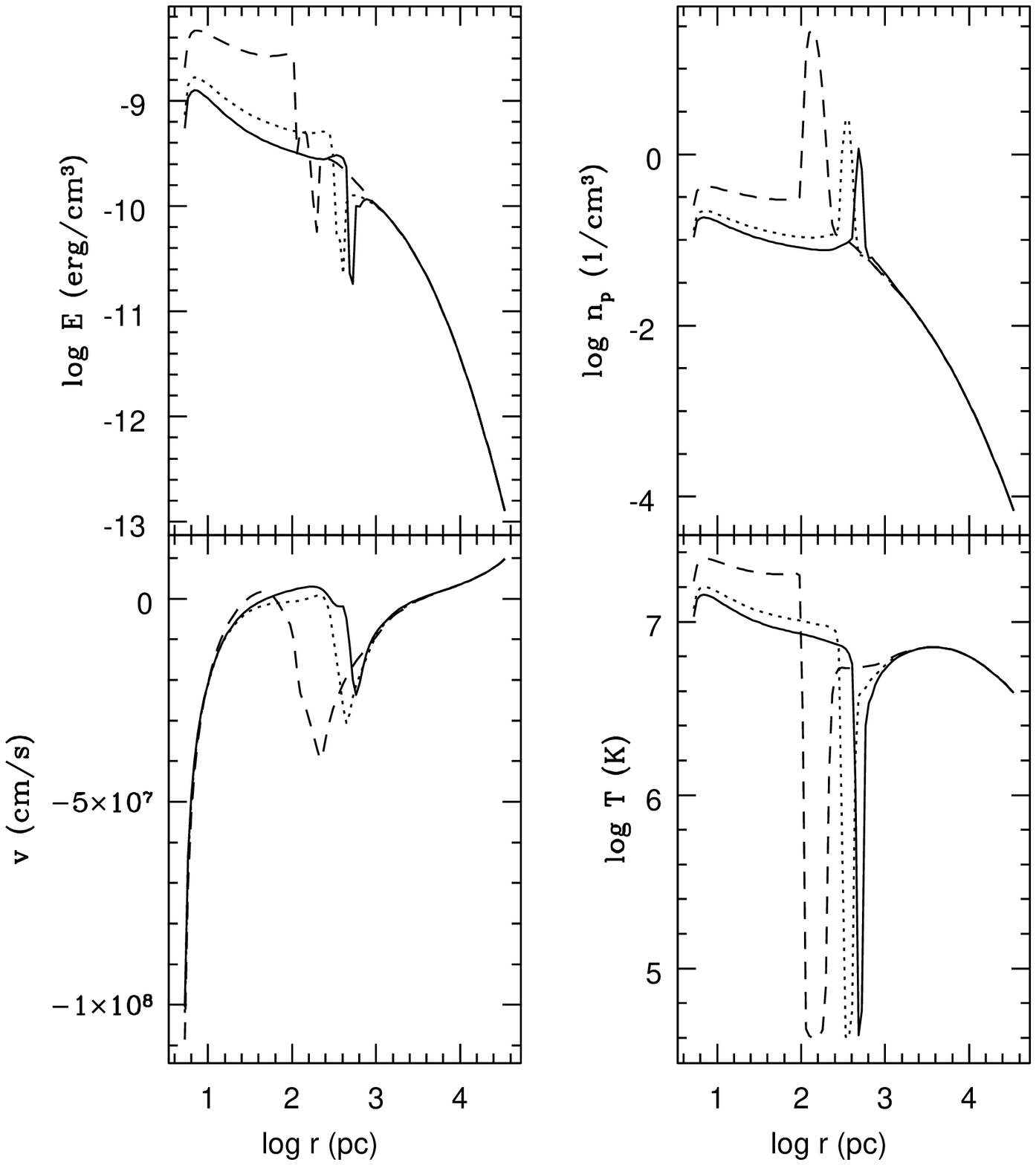}
\caption{Hydrodynamical quantities of the model immediately before 
the onset of the bursting phase ($2.554$ Gyr),  
separated by 1 Myr (in order, solid, dotted, dahsed). 
The cold shell is falling towards the galaxy center.}
\label{hydA}
\end{figure}

\begin{figure}
\includegraphics[angle=0.,scale=1]{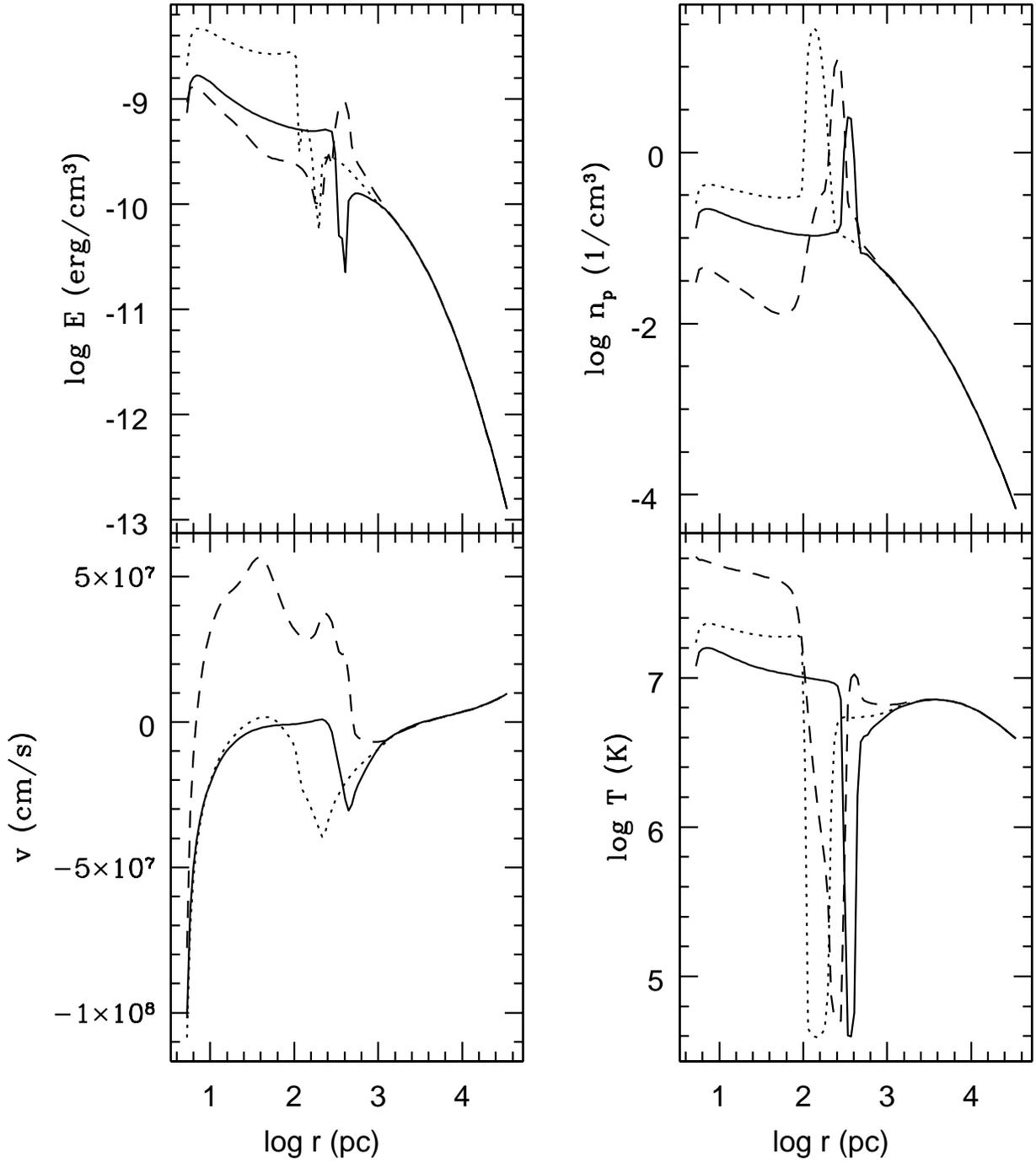}
\caption{The cold shell reached the center, and a shock wave is moving 
outward (dashed lines, note the positive velocities). 
Time interval is 1 Myr (in order, solid, dotted and dashed lines), 
and the first two snapshots are the two last snapshots in Fig.~\ref{hydA}.}
\label{hydB}
\end{figure}

\begin{figure}
\includegraphics[angle=0.,scale=1]{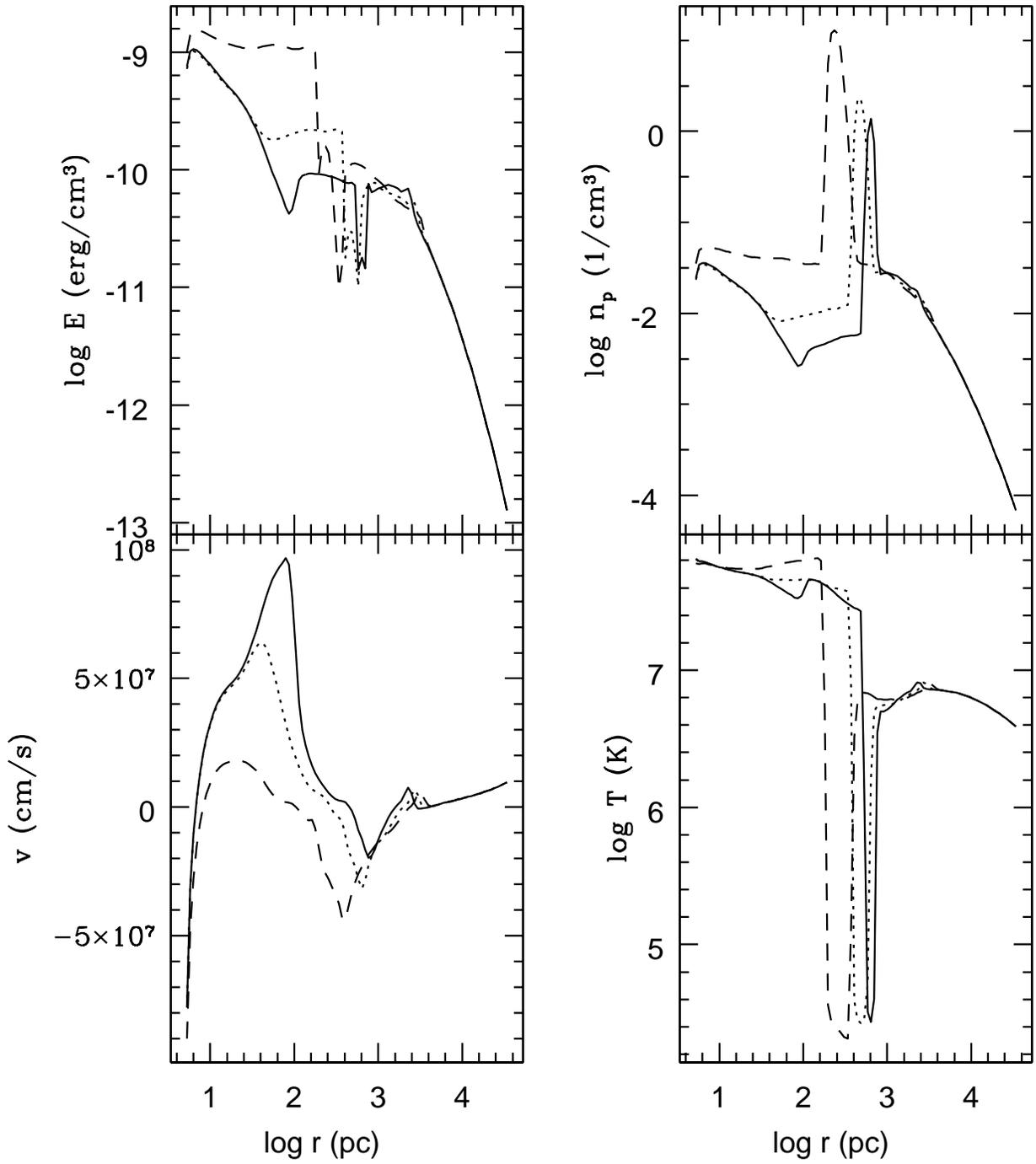}
\caption{The expanding flow produced a new cold shell, which starts to 
fall to the galactic center. Time interval is 1 Myr (in order, 
solid, dotted and dashed lines), while the time is now 2.561 Gyr.}
\label{hydC}
\end{figure}

\begin{figure}
\includegraphics[angle=0.,scale=1]{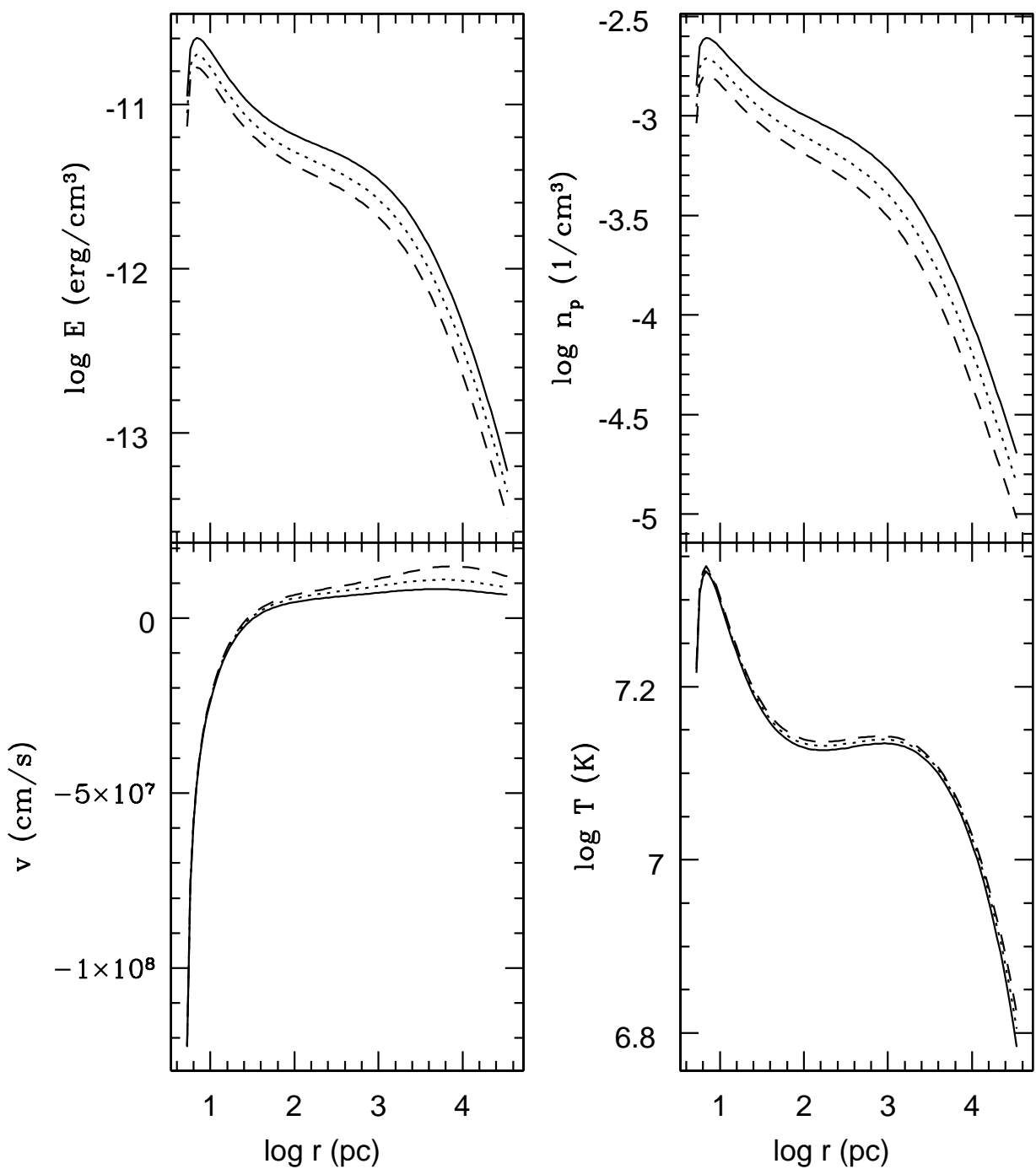}
\caption{Hydrodynamical quantities of the model at the end of the
simulation, separated by 0.5 Gyrs (in order, solid, dotted and dashed
lines).  The galaxy is in the low-luminosity, high-temperature and
low-density stationary accretion phase in the inner parts and, in its
outer parts a $\sim$100-200 km s$^{-1}$ wind is carrying out most of the
late recycled gas.}
\label{hydD}
\end{figure}

As was already found in CO01, the galaxy cooling catastrophe starts
with the formation of a {\it cold shell} placed around the galaxy core
radius: however, in the present models (as in those explored by
Pellegrini \& Ciotti 1998), the cooling catastrophe happens at
significantly earlier times than in CO01 models, both due to the
higher central stellar density and to the different time dependence
and amount of SNIa explosions.  The three main evolutionary phases of
the model are summarized in Figs.~\ref{hydA}-\ref{hydD}. In
particular, in Fig.~\ref{hydA} one can observe (with time separation
of 1 Myr), the evolution of the first cold shell falling to the galaxy
center, while in Fig.~\ref{hydB} (dashed line) the expanding material
due to the first burst (with velocities of the order of several
hundred km s$^{-1}$) is clearly visible. A particularly important
feature can be noticed in Fig.~\ref{hydC}, where a new cold and dense
shell of gas is formed as a consequence of the shock wave produced by
the burst. This shell moves (slowly) outwards, and then starts to fall
back at the center; in the shell, star formation rate reaches values
of $\sim 10^{-8}\Msun$ yr$^{-1}$ pc$^{-3}$.  We stress that this shell
is of a different origin compared to the first, and it is possibly
Rayleigh-Taylor unstable. This cycle of shell formation, central
burst, and expanding phase, repeats during all the bursting evolution,
along the lines described in detail in CO01.  Finally, when the
specific SNIa heating becomes dominant over the decline of fresh mass
input from evolving stars, the galaxy hosts a wind, and the accretion
becomes stationary without oscillations, and the central SMBH is
radiating at $\sim 10^{-5}\ledd$ (Hopkins, Narayan \& Hernquist
2006). The hydrodynamical quantities (separated now by a time interval
of 0.5 Gyrs) are shown in Fig.~\ref{hydD}.

\section{Conclusions and discussion}

In this paper we have addressed the effects of radiative feedback on
the gas flows in elliptical galaxies. The investigation is in the line
of previous exploratory papers (CO97, CO01, OC05), but now the input
physics and the galaxy models have been substantially improved. We
briefly recall here the main points on which our framework is
based. First of all, it is obvious that the recycled gas from dying
stars is an important source of fuel for the central SMBH, {\it even
in absence of external phenomena such as galaxy merging}, that are
often advocated as the way to induce QSO activity. It is also obvious
that the recycled gas, arising from stars in the inner several kpc of
the galaxy (assumed a giant elliptical), will necessarily drive a
classical radiative instability and a collapse towards the center of
metal rich gas. As a consequence, a star-burst must occur and also the
central SMBH must be fed. The details of how much is accreted on the
central SMBH vs. consumed in stars vs. ejected from the center by
energy input from the starburst and AGN are uncertain.  But order of
magnitude estimates would have the bulk going into stars or blown out
as a galactic wind, with a small amount going into the central SMBH.
In addition, since at the end of a major outburst an hot bubble
remains at the center, both processes shut themselves off, and it will
take a cooling time for the cycle to repeat. In other words,
relaxation oscillations are to be expected, but their detailed
character is uncertain.  Finally, order of magnitude estimates would
indicate that during the bursting phase the center would be optically
thick to dust, so one would observe a largely obscured starburst and
largely obscured AGN with most radiation in the far IR; as gas gets
consumed, the central sources become visible. Much of the AGN output
occurs during obscured phases: then there is a brief interval when one
sees a "normal" quasar, and finally one would see a low X-ray
luminosity and E+A spectrum galaxy, with A dominating in the central
several hundred pc for $10^{7-8}$ yrs.

The present paper attempts to illustrate the expectations described
above.  Overall, we have confirmed that radiative feedback from a
central SMBH has dramatic effects on galaxy evolution and on the mass
growth of the central SMBH itself, and we find that much of the
recycled gas falling towards the galaxy center during the accretion
events is consumed in central starbursts with a small fraction (of the
order of 1\% or less) accreted onto the central SMBH.  In particular,
in the presented model approximately half of the recycled mass from
evolving stars is expelled in the ICM, while the remaining fraction is
consumed in central starbursts.  Thus, the central starburst is an
important component in the physical modelling, since it regulates the
amount of gas available to be accreted onto the central SMBH: without
allowing for the (AGN feedback induced) central star formation the
SMBH would grow to be far more massive than seen in real galaxies.
While the details predicted by our simulations are uncertain, we do
not see how this sequence can be avoided in its qualitative features.
The input is standard physics plus the known radiative output from the
SMBH and stars.  Other processes that we do not include in the present
code may also be important. For example, the mechanical energy from
the AGN and cosmic rays from SNR in the starburst is surely important
(e.g., see Di Matteo et al. 2005), but we deliberately excluded them
from the present investigation to better assess the importance of
radiative feedback (both as heating and radiative pressure).

The main results of our simulations, also considering all the
simplifications in the treatment of physics, and of the geometry of
the code, nicely support the scenario depicted above. In particular,
{\it we showed how complicated the evolution of an isolated galaxy,
subject to internal evolution only, can be} (e.g., see Pierce et
al. 2007). From the observational point of view, we find that evidence
for starbursts should be common when looking at elliptical galaxies,
with the fraction showing an E+A spectrum increasing with redshift
(with preliminary evidence for high metallicity in the new stellar
spectra). Also, signs of star-formation in high-$z$ objects such as
those detected in UV ($GALEX$) and IR ($Spitzer$), or accompanying AGN
(X-ray), should be common (e.g., see Yan et al. 2006; Nesvadba et
al. 2006, Simoes Lopes et al. 2007), with IR luminosity peaks of $\sim
10^{46}$ erg s$^{-1}$. Due to the observational relevance of these
predictions, a more accurate modelization of the expected IR model
properties would be a natural extension of the present work (e.g.,
Chakrabarti et al. 2007).

We note that there is increasing evidence in the local universe of hot
gas disturbances on various galactic scales, likely the resultant from
recent nuclear activity (e.g., Forman et al. 2006).  For example,
$Chandra$ revealed two symmetric arm-like features across the center
of NGC4636 (Jones et al. 2002; O'Sullivan, Vrtilek, \& Kempner 2005),
accompanied by a temperature increase with respect to the surrounding
hot ISM; they were related to shock heating of the ISM, caused by a
recent nuclear outburst.  Other evidences include a hot filament in
the nuclear regions of NGC821 and NGC3377 (Fabbiano et al. 2004, Soria
et al. 2006); a "bar" feature, presumably due to a shock, at the
center of NGC4649 (Randall, Sarazin \& Irwin 2004); a nuclear outflow
in NGC4438 (Machacek, Jones, \& Forman 2004); an unusual temperature
profile finally is present in NGC 3411 (Vrtilek et al. 2006).

In this paper we presented just one out many models, to illustrate in
detail the global behavior of a typical solution.  Of course, the
present work suffers of some weak points, as: 1) we neglected the
modifications of the galaxy gravity field and velocity dispersion
profile due to the stellar mass losses, galactic wind, and star
formation (deepening the central potential). 2) The new stars are
placed in the galaxy where they form. 3) The additional effect of
mechanical feedback was not taken into account (e.g., see Begelman \&
Ruszkowski 2005). 4) Finally, the simulations are spherically
symmetric, so that Rayleigh-Taylor unstable configurations of the ISM,
and the formation of an accretion disk, cannot be followed.

Several lines of investigation will be studied in future papers, in
order to better test the results and to address specific points only
mentioned here. For example, it is natural to study in detail the
properties expected for the starburst population (such as spatial
distribution, spectral properties, etc.), and the X-ray properties of
the perturbed ISM, as a function of the combined effect of SNIa and
central feedback. Other obvious issues to be addressed are the effects
of environment, as for a cD galaxy in a cluster; this study will also
probe the impact of photoionization plus Compton heating on the ICM
(extending the preliminary investigation of Ciotti, Ostriker, \&
Pellegrini 2004, see also Pope et al. 2007).  Also, the modifications
of the galaxy inner structure and dynamics due to star formation are
important, in particular for the expected evolution of the Magorrian
and $\mbh -\sigma$ relations. A most relevant extension of the present
work would also be the study of radiative feedback by using
two-dimensional codes, in order to follow the evolution of unstable
features here found, such as the cold-shell phase, and to properly
describe axisymmetric accretion in the optically thick regime (e.g.,
Krolik 2007, Proga 2007).  We finally mention another observational
riddle that could be addressed with the present models, i.e., that of
the apparent ``underluminosity'' of SMBHs in the local universe (e.g.,
see Fabian \& Canizares 1988; Pellegrini 2005).  The present models
could also be used in a different way. In fact, when addressing galaxy
evolution over cosmological times, two main effects have not been
considered in the present treatment, namely galaxy formation itself
and the successive cosmological infall and merging events (e.g.,
Hopkins \& Hernquist 2006, Micic et al. 2007).  These aspects of
evolution are usually considered in cosmological simulations (e.g.,
see Springel, Di Matteo, \& Hernquist 2005; Naab et al. 2007, and
references therein), and so it would be interesting to add at the
center of galaxies in those simulations the treatment of feedback
physics here described.

\acknowledgments

We thank Annibale D'Ercole, Bruce Draine, Martin Elvis, Roberto Gilli,
Jeremy Goodman, Avi Loeb, Eve Ostriker, Silvia Pellegrini and Todd
Thompson for useful discussions, and the anonymous Referee for
important remarks. L.C. was partially supported by the grant CoFin2004
by Italian MIUR.

\appendix

\section{Numerical evaluation of lag integrals}

Time-lag integrals often appear in the code.  It is possible to
compute them numerically without much use of computer
memory\footnote{Note that in CO01 the exponential factors in front of
the integral in equation (B2) and inside the integral in equation (B3)
are two typos.}. In fact, let $f(t)$ a given function of time, and let
\beq
F(t)=\int_0^tf(t')\;e^{-{t-t'\over\tlag}}\;dt' 
\eeq
the quantity to be determined.  It is easy to prove that for $t_i\leq
t$ the exact identity
\beq
F(t)=F(t_i)e^{-{t-t_i\over\tlag}} + 
    \int_{t_i}^t f(t')\;e^{-{t-t'\over\tlag}}\;dt'
\eeq
holds, so that only the storage of values $F(t_i)$ and $f(t_i)$ is
necessary to evaluate $F(t_{i+1})$. Note that equation (A1) is the
solution of the differential equation
\beq
{d F\over dt}=f(t)-{F(t)\over\tlag},
\eeq
and this leads to an alternative way to compute $F(t)$ by using 
(for instance) a finite difference integration scheme.  In the integration
of the hydrodynamical equations the evaluation of the quantity $\Delta
F=\int_{t_{\rm i}}^{t_{\rm i+1}}F(t)\,dt$ over a time-step is also
often required. From equations (A1) and (A2), simple algebra proves
the exact relation
\beq
\Delta F=\tlag [F(t_i)-F(t_{i+1})]+ \tlag \int_{t_i}^{t_{i+1}}f(t')\;dt'.
\eeq
In the code, the function $f$ is defined as the linear interpolation
between the initial and final time over the time-step, so that the
integrals (A2) and (A4) can be explicitly calculated with negligible
computer time.

\clearpage


\begin{thebibliography}

\bibitem[]{} Begelman, M.C., \& Nath, B.B., 2005,
             \mnras, 361, 1387

\bibitem[]{} Begelman, M.C., \& Ruszkowski, M., 2005,
             Phil.Trans. of Roy.Soc., part A, 363, n.1828, 655

\bibitem[]{} Bertin, G., et al., 1994,
             A\&A, 292, 381

\bibitem[]{} Binney, J., 2001,
             in "Particles and Fields in Radio Galaxies Conference", 
             ASP Conference Proceedings, Robert A. Laing and 
             Katherine M. Blundell eds., vol. 250, p. 481

\bibitem[]{} Binney, J., \& Tabor, G., 1995,
             \mnras, 276, 663

\bibitem[]{} Burkert, A., \& Silk, J, 2001,
             \apj, 554, L151

\bibitem[]{} Cappellari, M., et al., 2006,
             \mnras, 366, 1126

\bibitem[]{} Cappellaro, E., Evans, R., \& Turatto, M., 1999,
             A\&A, 351, 459

\bibitem[]{} Cavaliere, A., \& Vittorini, V, 2002,
             \apj, 570, 114

\bibitem[]{} Cen, R., \& Ostriker, J.P., 2006,
             \apj, 650, 560

\bibitem[]{} Chakrabarti, S., Cox, T.J., Hernquist, L., Hopkins, P.F., 
             Robertson, B., \& Di Matteo, T., 2007,
             \apj, 658, 840

\bibitem[]{} Chandrasekhar, S., 1960,
             {\it Radiative Transfer}, (Dover, New York)

\bibitem[]{} Churazov, E., Sazonov, S., Sunyaev, R., Forman, W., Jones, C., \&
             B\"ohringer, H., 2005,
             \mnras, 363, L91

\bibitem[]{} Cimatti, A., et al., 2002,
             A\&A, 381, L68

\bibitem[]{} Ciotti, L., 1996,
             \apj, 471, 68

\bibitem[]{} Ciotti, L., \& Bertin, G., 1999,
             A\&A, 352, 447

\bibitem[]{} Ciotti, L., \& Ostriker, J.P., 1997,
             \apjl, 487, L105 (CO97)

\bibitem[]{} Ciotti, L., \& Ostriker, J.P., 2001, 
             \apj, 551, 131 (CO01)

\bibitem[]{} Ciotti, L., \& Pellegrini, S., 1992,
             \mnras, 255, 561

\bibitem[]{} Ciotti, L., Ostriker, J.P., \& Pellegrini, 2004, 
             In the Proceedings of the International Symposium 
             "Plasmas in the laboratory and in the universe: new 
             insights and new challenges", G. Bertin, D. Farina, 
             R. Pozzoli eds., AIPCS, vol.703, p.367

\bibitem[]{} Ciotti, L., D'Ercole, A., Pellegrini, S., \& Renzini, A., 1991, 
             \apj, 376, 380 (CDPR)

\bibitem[]{} Ciotti, L., Lanzoni, B., \&  Renzini, A., 1996,
             \mnras, 282, 1

\bibitem[]{} Cowie, L.L., Ostriker, J.P., \& Stark, A.A., 1978,
             \apj, 226, 1041

\bibitem[]{} Croton, D.J., et al., 2006,
             \mnras, 365, 11

\bibitem[]{} D'Ercole, A., Renzini, A., Ciotti, L.  \& Pellegrini, S., 1989,
             \apj, 341, L9

\bibitem[]{} Davies, R.I., Mueller S\'anchez, F., Genzel, R., Tacconi, L., 
             Hicks, E., Friedrich, S., \& Sternberg, A., 2007,
             preprint (arXiv:0704.1374)

\bibitem[]{} Di Matteo, T., Springel, V., \& Hernquist, L., 2005,
             Nature, 433, 604

\bibitem[]{} Douglas, N.G., et al., 2007,
             preprint (astro-ph/0703047)

\bibitem[]{} Dubinski, J., \& Carlberg, R.G., 1991,
             \apj, 378, 496

\bibitem[]{} Elvis, M., 2006,
             Mem.SaIt, 77, 573

\bibitem[]{} Fabbiano, G., Baldi, A., Pellegrini, S., Siemiginowska, A., 
             Elvis, M., Zezas, A., \& McDowell, J., 2004,
             \apj, 616, 730 

\bibitem[]{} Faber, S.M., et al., 1997,
             \aj, 114, 1771

\bibitem[]{} Fabian, A.C., 1999,
             \mnras, 308, L39

\bibitem[]{} Fabian, A.C., \& Canizares, C.R., 1988,
             Nature, 333, 829

\bibitem[]{} Fabian, A.C., Thomas, P.A., Fall, S.M., \& White III, R.E., 1986,
             \mnras, 221, 1049

\bibitem[]{} Fabian, A.C., Celotti, A., \& Erlund, M.C., 2006,
             \mnras, 373, L16

\bibitem[]{} Ferrarese, L., \& Merritt, D. 2000,
             \apj, 539, L9


\bibitem[]{} Forman, W., et al., 2006,
             preprint (astro-ph/0604583)

\bibitem[]{} Fukugita, M., Nakamura, O., Turner, E.L., Helmboldt, J., \&
             Nichol, R.C., 2004, 
             \apj, 601, L127

\bibitem[]{} Fukushige, T., \& Makino, J., 1997,
             \apj, 477, L9

\bibitem[]{} Gebhardt, K., et al., 2000,
             \apj, 539, L13

\bibitem[]{} Gilli, R., Comastri, A., \& Hasinger, G., 2007,
             A\&A, 463, 79

\bibitem[]{} Goodman, J., \& Tan, J.C., 2004,
             \apj, 608, 108

\bibitem[]{} Graham, A.W., \& Driver, S.P., 2005,
             PASA, 22, 118

\bibitem[]{} Graham, A.W., Erwin, P., Caon, N., \& Trujillo, I., 2003,
             Rev.Mex.A.A., 17, 196

\bibitem[]{} Granato, G.L., De Zotti, G., Silva, L., Bressan, A., \& 
             Danese, L., 2004,
             \apj, 600, 580

\bibitem[]{} Greggio, L., 2005,
             A\&A, 441, 1055

\bibitem[]{} Haiman, Z., Ciotti, L. \& Ostriker, J. P. 2004,
             \apj, 606, 204

\bibitem[]{} Hernquist, L., 1990,
             \apj, 356, 359

\bibitem[]{} Hopkins, P.F., \& Hernquist, L., 2006,
             \apjs, 166, 1

\bibitem[]{} Hopkins, P.F., Narayan, R., \& Hernquist, L., 2006,
             \apj, 643, 641

\bibitem[]{} Hopkins, P.F., Hernquist, L., Cox, T.J., Robertson, B., 
             Di Matteo, T., \& Springel, V., 2006,
             \apj, 639, 700

\bibitem[]{} Humphrey, P.J., Buote, D.A., Gastaldello, F., Zappacosta, L., 
             Bullock, J.S., Brighenti, F., Mathews, W.G., 2006,
             \apj, 646, 899

\bibitem[]{} Imanishi, M., Dudley, C.C., Maiolino, R., Maloney, P.R.,
             Nakagawa, T., \& Risaliti, G., 2007,
             (astro-ph/0702136)

\bibitem[]{} Jaffe, W., Ford, H.C., O'Connell, R.W., van den Bosch, F.C.,
             Ferrarese, L., 1994, 
             \aj, 108, 1567

\bibitem[]{} Jones, C., Forman, W., Vikhlinin, A., Markevitch, M., David, L.,
             Warmflash, A., Murray, S., \& Nulsen, P.E.J., 2002, 
             \apj, 567, L115


\bibitem[]{} King, I.R, 1972,
             \apj, 174, L123

\bibitem[]{} King, A.R., 2003, 
             \apj, 596, L27

\bibitem[]{} King, A.R., Pringle, J.E., \& Livio, M. 2007,
             preprint (astro-ph/0701803)

\bibitem[]{} Krolik, J. H., 2007,
             (astro-ph/0702396)

\bibitem[]{} Kr\"ugel, E., \& Tutukov, A.V., 1993,
             A\&A, 275, 416

\bibitem[]{} Lauer, T.R., et al., 2005,
             \aj, 129, 2138

\bibitem[]{} McLure, R.J., \& Dunlop, J.S., 2002,
             \mnras, 331, 795

\bibitem[]{} Machacek, M.E., Jones, C., \& Forman, W.R., 2004, 
             \apj, 610, 183

\bibitem[]{} Magorrian, J., et al. 1998,
             \aj, 115, 2285

\bibitem[]{} Mannucci, F., Della Valle, M., Panagia, N., Cappellaro, E., 
             Cresci, G., Maiolino, R., Petrosian, A., \& Turatto, M., 2005,
             A\&A, 433, 807

\bibitem[]{} Maraston, C., 2005, 
             \mnras, 362, 799

\bibitem[]{} Martinez-Sansigre, A., \&  Rawlings, S., 2007
             (astro-ph/0701143)

\bibitem[]{} Matteucci, F., Panagia, N., Pipino, A., Mannucci, F., Recchi, S., 
             \& Della Valle, M., 2006,
             \mnras, 372, 265

\bibitem[]{} Micic, M., Holley-Bockelmann, Sigurdsson, S., \& Abel, T., 2007,
             preprint (astro-ph/0703540)

\bibitem[]{} Murray, N., Quataert, E., \& Thompson, T.A., 2005,
             \apj, 618, 569

\bibitem[]{} Naab, T., Johansson, P.H., Ostriker, J.P., \& Efstathiou, G., 
             2007, \apj, in press (astro-ph/0512235) 

\bibitem[]{} Narayan, R. \& Yi, I., 1994
             \apj, 428, L13

\bibitem[]{} Navarro, J.F., Frenk, C.S., \& White, S.D.M., 1997,
             \apj, 490, 493

\bibitem[]{} Nayakshin, S., \& Sunyaev, R., 2005,
             \mnras, 364, L23

\bibitem[]{} Nayakshin, S., Dehnen, W., Cuadra, J., \& Genzel, R., 2006,
             \mnras, 366, 1410

\bibitem[]{} Neill, J.D., et al., 2007, preprint
             (astro-ph/0701161)

\bibitem[]{} Nesvadba, N.P.H., Lehnert, M.D., Eisenhauer, F., Gilbert, A., 
             Tecza, M., \& Abuter, R., 2006,
             \apj, 650, 693

\bibitem[]{} Nolan, L.A., Raychaudhury, S., \& Kab\'an, A., 2007,
             \mnras, 375, 381

\bibitem[]{} O'Sullivan, E., Ponman, T.J., \& Collins, R.S., 2003,
             \mnras, 340, 1375   

\bibitem[]{} O'Sullivan, E., Vrtilek, J.M., \& Kempner, J.C., 2005,
             \apj, 624, L77

\bibitem[]{} O'Sullivan, E., Vrtilek, J.M., Harris, D.E., \& Ponman, T.J., 
             2006, (astro-ph/0612287)

\bibitem[]{} Omma, H., Binney, J., Bryan, G., \& Slyz, A., 2004, 
             \mnras, 348, 1105

\bibitem[]{} Ostriker, J.P., \& Ciotti, L., 2005,
             Phil.Trans. of Roy.Soc., part A, 363, n.1828, 667 (OC05)

\bibitem[]{} Ostriker, J.P.,  Weaver, R., Yahil, A., \& McCray, R., 1976,
             \apjl, 208, L61

\bibitem[]{} Pellegrini, S., 2005,
             \apj, 624, 155

\bibitem[]{} Pellegrini, S., \& Ciotti, L., 1998,
             A\&A, 333, 433

\bibitem[]{} Pellegrini, S., \& Ciotti, L., 2006,
             \mnras, 370, 1797

\bibitem[]{} Peterson, J.R., \& Fabian, A.C., 2006,
             Phys.Rep., 427, 1

\bibitem[]{} Pierce, C.M., et al., 2007,
             \apj, 660, L19

\bibitem[]{} Pope, A., et al., 2006,
             \mnras, 370, 1185

\bibitem[]{} Pope, E.C.D., Pavlosvski, G., Kaiser, C.R., \& Fangohr, H., 2007,
             (astro-ph/0702683)

\bibitem[]{} Proga, D., 2007,
             (astro-ph/0702582)

\bibitem[]{} Springel, V., Di Matteo, T., \& Hernquist, L., 2005,
             \mnras, 361, 776

\bibitem[]{} Randall, S.W., Sarazin, C.L., \& Irwin, J.A., 2004,
             \apj, 600, 729

\bibitem[]{} Recchi, S., D'Ercole, A., \& Ciotti, L., 2000,
             \apj, 533, 799

\bibitem[]{} Renzini, A., \& Ciotti, L., 1993,
             \apj, 416, L49

\bibitem[]{} Renzini, A., Ciotti, L., D'Ercole, A. \& Pellegrini, S., 1993,
             \apj, 419, 52

\bibitem[]{} Reuland, M., et al., 2007,
             (astro-ph/0702753)

\bibitem[]{} Riciputi, A. Lanzoni, B., Bonoli, S., \&  Ciotti, L., 2005,
             A\&A, 443, 133

\bibitem[]{} Roberts, M.S., Hogg, D.E., Bregman, J.N., Forman, W.R., \& 
             Jones, C., 1991, 
             \apjs, 75, 751

\bibitem[]{}  Rodighiero, G., et al. 2007
             (astro-ph/0701178)

\bibitem[]{} Russell, P.A., Ponman, T.J., \& Sanderson, A.J.R., 2007,
             (astro-ph/0703010)

\bibitem[]{} Saglia, R.P., et al., 1993, 
             A\&A, 403, 567
 
\bibitem[]{} Sazonov, S.Yu., Ostriker, J.P., \& Sunyaev, R., 2004,
             \mnras, 347, 144

\bibitem[]{} Sazonov, S.Yu., Ostriker, J.P., Ciotti, L., \& Sunyaev, R.A., 
             2005, \mnras, 358, 168

\bibitem[]{} Sazonov, S.Yu., Revnivtsev, M., Krivonos, R., Churazov, E., \& 
             Sunyaev, R.A., 2007, 
             A\&A, 462, 57

\bibitem[]{} Sersic, J.L., 1968,
             Atlas de galaxias australes. Observatorio Astronomico, Cordoba

\bibitem[]{} Silk, J., \& Rees, M.J., 1998,
             A\&A, 331, L1

\bibitem[]{} Simoes Lopes, R.D., Storchi-Bergmann, T., de Fatima Saraiva, M., 
             \& Martini, P., 2007,
             \apj, 655, 718

\bibitem[]{} Soltan, A., 1982,
             \mnras, 200, 115

\bibitem[]{} Soria, R., Fabbiano, G., Graham, A.W., Baldi, A., Elvis, M., 
             Jerjen, H., Pellegrini, S., \& Siemiginowska, A., 2006,
             \apj, 640, 126

\bibitem[]{} Spergel, D.N., et al., 2006, 
             preprint (astro-ph/0603449)

\bibitem[]{} Tan, J.C., \& Blackman, E.G., 2005,
             \mnras, 362, 983


\bibitem[]{} Thompson, T.A., Quataert, E., \& Murray, N., 2005,
             \apj, 630, 167

\bibitem[]{} Thompson, T.A., Quataert, E., \& Waxman, E., 2006,
             preprint (astro-ph/0606665)

\bibitem[]{} Treu, T., Koopmans, L.V., Bolton, A.S., Burles, S., \& 
             Moustakas, L.A., 2006, 
             \apj, 640, 662

\bibitem[]{} Weinberg, D.H., Hernquist, L., \& Katz, N., 2002, 
             \apj, 571, 15 

\bibitem[]{} Wyithe, J.S.B., \& Loeb, A., 2003,
             \apj, 595, 614

\bibitem[]{} Xu, H., et al., 2002,
             \apj, 579, 600

\bibitem[]{} Yan, R., et al., 2006,
             AAS 209, 181.05

\bibitem[]{} Yu, Q., Tremaine, S., 2002, 
             \mnras, 335, 965

\end{thebibliography}
\end{document}